\documentclass[aps,showpacs,nofootinbib,superscriptaddress,prd]{revtex4}
\usepackage{graphicx}
\usepackage{amssymb}
\usepackage{amsmath}
\usepackage{color}
\usepackage{multirow}

\newcommand\f[2]{\frac{#1}{#2}}
\newcommand\as{\alpha_s}
\def\nn{\nonumber}
\def\ito{\leftarrow}

\begin{document}

\title{Improved Resummation Prediction on Higgs Boson Production at Hadron Colliders}

\author{Jian Wang}
\affiliation{Department of Physics and State Key Laboratory
of Nuclear Physics and Technology, Peking University,
Beijing 100871, China}

\author{Chong Sheng Li}
\email{csli@pku.edu.cn}
\affiliation{Department of Physics and State Key Laboratory
of Nuclear Physics and Technology, Peking University,
Beijing 100871, China}
\affiliation{Center for High Energy Physics,
Peking University, Beijing 100871, China}

\author{Hai Tao Li}
\affiliation{Department of Physics and State Key Laboratory
of Nuclear Physics and Technology, Peking University,
Beijing 100871, China}

\author{Zhao Li}
\email{zhaoli@pa.msu.edu}
\affiliation{Department of Physics \& Astronomy,
Michigan State University, E. Lansing, MI 48824, USA}

\author{C.-P. Yuan}
\email{yuan@pa.msu.edu}
\affiliation{Center for High Energy Physics,
Peking University, Beijing 100871, China}
\affiliation{Department of Physics \& Astronomy,
Michigan State University, E. Lansing, MI 48824, USA}

\pacs{12.38.Cy,12.38.Qk,14.80.Bn}

\begin{abstract}
We improve the resummation calculations in the ResBos program for
the Higgs boson production via gluon-gluon fusion by including the
NNLO Wilson coefficient functions and G-functions.
The improvement leads to increasing the total cross section predictions of the new ResBos program, dubbed as ResBos2,
for Higgs boson production
by about $8\%$ and $6\%$ at the Tevatron and the LHC, respectively, as compared to the old ResBos program.
Furthermore, the improved predictions are compared with those from the programs HNNLO and HqT2.
We find that they agree well for the total cross sections
but differ slightly for the transverse momentum $Q_T$ distributions.
With ResBos2, we present the distributions of the two variables $a_T$  and $\phi^{*}$,
which can have better experimental resolutions than $Q_T$,
for the process of Higgs boson decaying into a photon pair.
Theoretical uncertainties of the ResBos2 predictions are also discussed.

\end{abstract}

\maketitle

\section{Introduction}
In the standard model (SM), the Higgs boson is responsible for the origin of electroweak symmetry  breaking
and the generation of elementary particle masses.
In the past decades, the Fermilab Tevatron has devoted great effort to searching
for Higgs boson in the SM, whose mass is not theoretically precisely predicted.
Up to now, the Tevatron has excluded a wide range of the Higgs boson mass.
With the data corresponding to an integrated luminosity of 10 fb$^{-1}$,
the CDF and D\O~ collaborations found about $2\sigma$ excess in the mass
region $115$ GeV $<m_H<135$ GeV \cite{CDFD0Higgs}.

At the CERN Large Hadron Collider (LHC),
the ATLAS \cite{ATLAS:2012ae} and CMS \cite{Chatrchyan:2012tx}  collaborations
found an about $3\sigma$ excess for the Higgs boson mass in the range $124$ GeV $<m_H<$ $126$ GeV
using the data collected  in 2011 at a center-of-mass (CM) energy of 7 TeV.
Most recently, as a milestone of particle physics,
a new resonance with a mass of about 125 GeV,
a Higgs-like particle, has been observed at a $5\sigma$ level \cite{ATLAS:2012gk,CMS:2012gu}
after the data collected at 8 TeV are taken into account.
In the future, when more data are accumulated, it is promising that LHC can tell
whether this is the SM Higgs boson or something else.

The SM Higgs boson is produced at the LHC  mainly through the gluon-gluon fusion process.
At the lowest order, the Higgs boson couples with gluons via a top quark loop \cite{PhysRevLett.40.692},
and the higher order corrections involve multi-loop diagrams.
The QCD next-to-leading order (NLO) correction for this process
has been investigated in both the infinite 
\cite{Dawson:1990zj,Djouadi:1991tka}
and finite \cite{Graudenz:1992pv,Spira:1995rr} top quark mass limits,
and can enhance the total cross section by more than $80\%$
for a 125 GeV Higgs boson produced at the LHC with a CM energy of 7 TeV.
The QCD next-to-next-to-leading order (NNLO) correction for
the inclusive total cross section has been obtained
in the infinite top quark mass limit \cite{Harlander:2002wh,Anastasiou:2002yz,Ravindran:2003um},
and increases the NLO cross section by about $25\%$.
The validity of taking this infinite top quark mass limit at NNLO has been studied
in Refs. \cite{Marzani:2008az, Harlander:2009mq, Pak:2009bx}.
Furthermore, the soft gluon contributions
have been resummed up to next-to-next-to-leading logarithm (NNLL)
\cite{Catani:2003zt},
leading to an additional increment of cross section by about $7\%$ at the LHC.
The next-to-next-to-next-to-leading logarithm (NNNLL) resummation has also been studied
in Refs. \cite{Moch:2005ky,Laenen:2005uz,Idilbi:2005ni}.
Recently, the contributions in the form of $(C_A \pi \alpha_s)^n$,
where $\alpha_s$ is the strong coupling constant
and $C_A(=3)$ is the Casimir operator for the adjoint representation of SU(3) group,
have been resummed to all order in the soft-collinear effective theory
\cite{Ahrens:2008qu,Ahrens:2008nc}.
Besides the total cross section,
a precise prediction of the transverse momentum distribution is
of crucial importance to understand the properties of the SM Higgs boson,
which has been investigated at the QCD NLO
\footnote{In this paper, for the transverse momentum distribution of Higgs boson production,
 the QCD NLO results are referred to as those up to $\mathcal{O}({\alpha_s^4})$.
}
\cite{Kauffman:1996ix,deFlorian:1999zd,Ravindran:2002dc,
Glosser:2002gm,Anastasiou:2005qj,Catani:2007vq,Grazzini:2008tf}.
Moreover, contributions at  $\mathcal{O}({\alpha_s^5})$ to the production of Higgs boson
associated with two jets have also been accomplished in Ref. \cite{Campbell:2006xx}.
Nonetheless, in order to obtain reliable theoretical prediction in the low transverse momentum region,
the logarithmically-enhanced contributions due to soft gluon emission have been resummed up to NNLL
\cite{Yuan:1991we,Balazs:2000wv,Balazs:2000sz,Berger:2002ut,Kulesza:2003wn,
Bozzi:2003jy,Bozzi:2005wk,Bozzi:2007pn,Cao:2007du,Cao:2009md,deFlorian:2011xf,Becher:2012qa}.

In this paper, we show the improvement on the transverse momentum resummation calculations
in the ResBos program \cite{Balazs:1997xd},
in which the large logarithms of the form ${\rm ln}^n(m_H^2/Q_T^2)$ have been resummed to all orders in $\as$.
The improved ResBos program includes the Wilson coefficient function up to NNLO
and the Sudakov exponent with  NNLL accuracy,
and is matched to the $\mathcal{O}(\alpha_s^4)$ fixed-order prediction in the large transverse momentum region.
Hereafter, we refer to the improved version of ResBos as ResBos2 \cite{ResBos2}.
We shall compare the ResBos2 predictions on the total cross section and transverse momentum distribution
with those from the parton level Monte Carlo programs HNNLO
\cite{Catani:2007vq,Grazzini:2008tf}
and HqT2  \cite{Bozzi:2003jy,Bozzi:2005wk,Bozzi:2007pn,deFlorian:2011xf}.
The HNNLO program gives prediction exactly at NNLO.
The HqT2 prediction includes resummation effects up to NNLL order,
and is matched to NLO results at large transverse momentum region.
Though both HqT2 and ResBos2 calculations are performed at the same NNLL+NLO order
in predicting the transverse momentum distribution of the Higgs boson,
the integrated rate of the HqT2 calculations is required to reproduce the QCD NNLO total cross section,
while there is no such requirement in ResBos2 calculations.
The difference in their numerical predictions will be discussed in detail in Sec. \ref{sec:nume}.

In the transverse momentum resummation scheme, the values of non-perturbative coefficients,
needed for predicting the low transverse momentum distribution,
are obtained by fitting to the experiment data of Drell-Yan processes.
The non-perturbative coefficients for $gg\rightarrow H$ are assumed to be scaled by the ratio $C_A/C_F$
from those for Drell-Yan processes,
where $C_F(=4/3)$ is the Casimir operator for the fundamental representation of SU(3) group.
However, the non-perturbative physics may not follow this assumption,
so when we provide the soft gluon resummation prediction on $gg\rightarrow H$,
the theoretical uncertainty due to the dependence on non-perturbative coefficients
is also investigated in Sec. \ref{sec:nume}.

This paper is organized as follows.
In Sec. \ref{sec:form}, we briefly review the formula of transverse momentum resummation for Higgs boson production,
and give explicit expressions of NNLO Wilson coefficient functions included in ResBos2.
In Sec. \ref{sec:nume}, we present the numerical results for the total cross section and the transverse
momentum distributions  given by ResBos2, and compare them with the HNNLO and HqT2 predictions.
We conclude in Sec. \ref{sec:conc}.

\section{Resummation Formula}
\label{sec:form}

In this section we briefly review the Collins-Soper-Sterman (CSS)
\cite{Collins:1981uk,Collins:1981va,Collins:1984kg}
formula of the soft gluon resummation for Higgs boson production and decay \cite{Cao:2009md}
used in the ResBos program \cite{Balazs:1997xd}.
By using the narrow width approximation, the Higgs boson production process can be factorized from its subsequent decay.
The resummation formula for the inclusive differential cross section of the $gg\to H$ process can be written as
\cite{Balazs:2000wv,Cao:2009md}
\begin{align}
 \label{eq:resum_formalism}
 & \frac{ d\sigma(gg\to HX ) }{ dQ^2 dQ_T^2 dy }
  = \kappa \sigma_0 \frac{Q^2}{S} \frac{Q^2 \Gamma_H/m_H}{(Q^2-m_H^2)^2+(Q^2\Gamma_H/m_H)^2}
\\
 & \times \Biggl\{ \frac{1}{(2\pi)^2} \int d^2b e^{iQ_T\cdot b}
\widetilde{W}_{gg}(b_*,Q,x_1,x_2,C_{1,2,3})
\widetilde{W}_{gg}^{NP}(b,Q,x_1,x_2)+Y(Q_T,Q,x_1,x_2,C_4)\Biggr\},
\nonumber
 \end{align}
where $S$ is the square of the CM energy;
$Q$, $Q_{T}$, $y$, and $\Gamma_H$ are the invariant mass, transverse momentum, rapidity,
and total decay width of the Higgs boson, respectively.
In Eq.~(\ref{eq:resum_formalism}), the quantity $\sigma_0$, defined as $\sigma^{\rm LO}(\infty, 0,0)$,
arises as an overall factor in the infinite top quark mass limit. It is given by
\begin{equation}
  \label{eq:sigma0}
\sigma_0 = \frac{\sqrt{2} G_F \alpha_s^2}{576 \pi},
\end{equation}
where $G_F$ is the Fermi constant and $\alpha_s$ is evaluated at the hard scale $C_2 Q$.
Furthermore, we have multiplied $\sigma_0$ by an additional factor
\begin{equation}
\label{eq:kappa}
\kappa = \frac{\sigma^{\rm LO}(m_t,m_b,m_c)}{\sigma^{\rm LO}(\infty,0,0)},
 \end{equation}
in order to take into account the mass effects of the top, bottom, and charm quarks at LO.
It has been shown that multiplying the NLO cross section in the infinite top quark mass limit
by the factor $\kappa$ is a good approximation to the full mass-dependent NLO cross section
over a wide range of the Higgs boson mass \cite{Kramer:1996iq}.
The total cross section can be obtained by integrating the result in Eq. (\ref{eq:resum_formalism})
over the transverse momentum and the rapidity.

In Eq. (\ref{eq:resum_formalism}), the term containing $\widetilde{W}_{gg}$ dominates at small $Q_T$,
and behaves as $Q_{T}^{-2}$ times a series of $\ln^n{(Q^2/Q_T^2)}$.
The explicit form of $\widetilde{W}_{gg}$ can be expressed as
\begin{equation}
\widetilde{W}_{gg}(b,Q,x_{1},x_{2},C_{1,2,3}) =
e^{-S(b,Q,C_1,C_2)}\sum_{a,b=q,\bar{q},g}
\bigl(C_{ga}\otimes f_a\bigr)(x_1)\bigl(C_{gb}\otimes f_b\bigr)(x_2),
\end{equation}
where the Sudakov exponent is given by
\begin{eqnarray}
\label{sudakov}
S(b,Q,C_1,C_2)& = &\int_{C_1^2/b^2}^{C_2^2Q^2}
\frac{d\bar{\mu}}{\bar{\mu}^2}\left[A(\alpha_s(\bar{\mu}),C_1)
\ln\left(\frac{C_2^2 Q^2}{\bar{\mu}^2}\right)+B(\alpha_s(\bar{\mu}),C_1,C_2)\right]\,.
\end{eqnarray}
The coefficients $A$ and $B$ and the Wilson coefficient functions $C_{ga}$
can be expanded as a power series in $\alpha_s$:
\begin{eqnarray}
A(\alpha_s(\bar{\mu}),C_1)& = &
\sum_{n=1}^\infty \left(\frac{\alpha_s(\bar{\mu})}{\pi}\right)^n A^{(n)}(C_1),\label{abfunction}
\\
B(\alpha_s(\bar{\mu}),C_1,C_2)&=&
\sum_{n=1}^\infty \left(\frac{\alpha_s(\bar{\mu})}{\pi}\right)^n B^{(n)}(C_1,C_2),
\end{eqnarray}
and
\begin{eqnarray}
C_{ga}(z,b,\mu,C_1,C_2)& = &
\sum_{n=0}^\infty \left(\frac{\alpha_s(\mu)}{\pi}\right)^n
C_{ga}^{(n)}(z,b,\mu,\frac{C_1}{C_2}).
\label{cfunction}
\end{eqnarray}
Here, the constants $C_1$ and $C_2$ are included in order to adjust the scales at which the coefficients $A$ and $B$ are evaluated,
so that the higher-order corrections are moderately small  \cite{Collins:1984kg}.
The renormalization scale $\mu$ in $C_{ga}$ is, in principle, arbitrary and set equal to $C_3/b$ in the ResBos program.
The constant $C_4$ in the non-singular part $Y$ of Eq. (\ref{eq:resum_formalism}) sets the scale of the fixed-order perturbative calculation which is of the order of the Higgs boson invariant mass $Q$.
Unless specified otherwise, we use the canonical choices for the renormalization constants, i.e.,
$
 C_1=C_3=b_0=2e^{-\gamma_E},  C_2=C_4=1,
$
in our numerical calculations.
The perturbation expansion quantities in Eqs. (\ref{abfunction})-(\ref{cfunction})
can be extracted order-by-order from fixed-order calculations.
In the old version of the ResBos program, we have included $A^{(1, 2, 3)}$,  $B^{(1, 2)}$ and $C^{(0, 1)}$,
whose analytical expressions are given in Ref. \cite{Cao:2009md}.

In this paper, we improve the ResBos calculations by including the NNLO Wilson coefficient functions,
which can be readily derived from ${\cal H}_{gg \ito ab}^{H (2)}$, as given in Eqs. (23) and (24)
of Ref. \cite{Catani:2011kr}.
As shown below, the explicit expressions of the NNLO Wilson coefficient
functions are
\begin{align}
C_{gq}^{(2)}(z)&=
C_A C_F
\left\{
   {\rm Li}_2(z)
   \left[
      \frac{z^2}{3}-\frac{z}{2}-\frac{11}{6 z}+
      \left(
         \frac{3 z}{4}+\frac{3}{2 z}-\frac{1}{2}
      \right)
      \ln(z)+2
   \right]
   -\frac{3 \left[(z+1)^2+1\right]}{4 z} {\rm Li}_3(-z)
                    \right. \nonumber \displaybreak[0] \\ & \left.
   +\left(-\frac{5 z}{4}-\frac{5}{2 z}+\frac{1}{2}\right)
   {\rm Li}_3(z)-\frac{\left[(1+z)^2+1\right] }{2 z}{\rm Li}_3\left(\frac{1}{z+1}\right)+
   {\rm Li}_2(-z)
   \left(\frac{z}{4}+\frac{(1+z)^2+1}{4 z} \ln(z)\right)
                    \right. \nonumber \displaybreak[0] \\ & \left.
   +\left(\frac{38}{27}-\frac{\pi ^2}{18}\right) z^2+
   \frac{-5z^2-22 (1-z)+6 \left[(1-z)^2+1\right] \ln (z)}{48 z} \ln^2(1-z)+
   \frac{1}{48} \left(8 z^2+9 z+36\right) \ln^2(z)
                    \right. \nonumber \displaybreak[0] \\ & \left.
   +\frac{\left[(1+z)^2+1\right] \left[3 \ln^2(z)-\pi ^2\right]-6 z^2 \ln(z)}{24 z} \ln(1+z)+
   \left(-\frac{11 z^2}{9}+\frac{z}{12}-\frac{1}{z}-\frac{107}{24}\right) \ln(z)
                    \right. \nonumber \displaybreak[0] \\ & \left.
   +\frac{-43 z^2+6 \left(4 z^3-9 z^2+24 z-22\right) \ln(z)+152 z+9
       \left[(1-z)^2+1\right] \ln^2(z)-152 }{72 z}\ln(1-z)
                    \right. \nonumber \displaybreak[0] \\ & \left.
   +\frac{1}{z}\left(4 \zeta (3)-\frac{503}{54}+\frac{11 \pi ^2}{36}\right)+
   z \left(2 \zeta(3)-\frac{133}{108}+\frac{\pi ^2}{3}\right)-
   \frac{(1-z)^2+1}{24 z} \ln^3(1-z)-
   \frac{1}{12} \left(\frac{z}{2}+1\right) \ln^3(z)
                    \right. \nonumber \displaybreak[0] \\ & \left.
   +\frac{(1+z)^2+1}{12 z} \ln^3(z+1)-
   \frac{5 \zeta_3}{2}-\frac{\pi ^2}{3}+\frac{1007}{108}
\right\}
-\frac{C_F z}{8} \left[\left(5+\pi ^2\right) C_A-3 C_F\right]+
C_A C_F \frac{2(1-z)+(z+1) \ln(z)}{z}
                    \nonumber \displaybreak[0] \\ &
+C_F^2
\left\{
   -\frac{13 z}{16}+\frac{1}{12} \left(\frac{z}{2}+\frac{1}{z}-1\right) \ln^3(1-z)+
   \frac{1}{48} (2-z) \ln^3(z)+\frac{1}{16}
   \left(z+\frac{6}{z}-6\right) \ln ^2(1-z)
                    \right. \nonumber \displaybreak[0] \\ & \left.
   -\frac{1}{32} (3 z+4) \ln^2(z)+
   \left(\frac{5 z}{8}+\frac{2}{z}-2\right) \ln(1-z)+
   \frac{5}{16}(z-3) \ln(z)+\frac{5}{8}
\right\}
                    \nonumber \displaybreak[0] \\ &
+C_F n_f
\left\{
   \frac{13 z}{108}+\frac{14}{27 z}+
   \frac{(1-z)^2+1}{24 z} \ln^2(1-z)+
   \frac{1}{18} \left(z+\frac{5}{z}-5\right) \ln(1-z)-
   \frac{14}{27}
\right\}
                    \nonumber \displaybreak[0] \\ &
-\frac{\left(z^2+2 z+2\right) \ln(z)-2 z^2}{z}\ln(z)\ln(1+z),\label{eq:cgq}
\end{align}
and
\begin{align}
2C_{gg}^{(2)}(z)&=
\delta (z-1)
\left\{
   C_A^2
   \left[
      \frac{7}{8} \ln\left(\frac{m_H^2}{m_t^2}\right)-\frac{55 \zeta_3}{18}+\frac{13 \pi ^4}{144}+
      \frac{157 \pi^2}{72}+\frac{3187}{288}
   \right]
   +C_A C_F \left[-\frac{11}{8}\ln\left(\frac{m_H^2}{m_t^2}\right)-\frac{3 \pi ^2}{4}-\frac{145}{24}\right]
                    \right. \nonumber \displaybreak[0] \\ & \left.
   -C_A n_f \left(\frac{4 \zeta_3}{9}+\frac{287}{144}+\frac{5 \pi ^2}{36}\right)-
   \frac{5 C_A}{96}+\frac{9 C_F^2}{4}+
   C_F n_f \left[\frac{1}{2} \ln\left(\frac{m_H^2}{m_t^2}\right)+\zeta_3-\frac{41}{24}\right]-
   \frac{C_F}{12}
\right\}
                    \nonumber \displaybreak[0] \\ &
+C_A^2 \left\{
   \frac{(1-z) \left(11 z^2-z+11\right)}{3 z}{\rm Li}_2(1-z)+
   \frac{\left(z^2+z+1\right)^2}{z (z+1)} \left[2 {\rm Li}_3\left(\frac{z}{z+1}\right)-{\rm Li}_3(-z)\right]
                    \right. \nonumber \displaybreak[0] \\ & \left.
   +\frac{\left(z^2+z+1\right)^2}{z (z+1)} \left[{\rm Li}_2(-z) \ln(z)-\frac{1}{3}
   \ln^3(z+1)+\frac{1}{6} \pi ^2 \ln(z+1)\right]
   -\frac{z^5-5 z^4+z^3+5 z^2-z+5}{z (1-z^2)}[{\rm Li}_3(z)-\zeta_3]
                    \right. \nonumber \displaybreak[0] \\ & \left.
   +\frac{z^5-3 z^4+z^3+3 z^2-z+3}{z (1-z^2)} {\rm Li}_2(z) \ln(z)
   +\frac{835 z^2}{54}-\frac{\left(-z^2+z+1\right)^2}{6 (1-z^2)}\ln^3(z)
                    \right. \nonumber \displaybreak[0] \\ & \left.
   +\left[
      \frac{1}{24} \left(44 z^2-11 z+25\right)+
      \frac{\left(z^2-z+1\right)^2}{2 z (1-z)} \ln(1-z)-
      \frac{\left(z^2+z+1\right)^2}{2 z (1+z)} \ln(1+z)
   \right] \ln^2(z)
                    \right. \nonumber \displaybreak[0] \\ & \left.
   +\left[
      \frac{\left(z^2-z+1\right)^2}{2 z (1-z)} \ln^2(1-z)+
      \frac{\left(z^2+z+1\right)^2}{z (1+z)} \ln^2(1+z)+
      \frac{-536 z^3-149 z^2-773 z-72}{72 z}
   \right] \ln(z)+
                    \right. \nonumber \displaybreak[0] \\ & \left.
   \frac{\left(-12 z^4-10 z^3-22 z^2-17 z+2\right) \zeta (3)}{2 z (z+1)}-
   \frac{380 z}{27}-\frac{449}{27 z}+\frac{1}{12} z \log (1-z)+\frac{517}{27}
\right\}
                    \nonumber \displaybreak[0] \\ &
+C_A^2 \frac{-2 z+(z+1) \ln(z)+2}{z}-
\frac{1}{16} \left[\left(5+\pi ^2\right) C_A-3 C_F\right]^2 \delta (1-z)
+\left(\frac{1}{1-z}\right)_+ \left[
   C_A^2 \left(\frac{7 \zeta_3}{2}-\frac{101}{27}\right)+
   \frac{14 C_A n_f}{27}
\right]
                    \nonumber \displaybreak[0] \\ &
+C_A n_f \left[
   -\frac{139 z^2}{108}+\frac{55 z}{54}+\frac{121}{108 z}+\frac{1+z}{12}\ln^2(z)-
   \frac{1}{12} z \ln(1-z)+\frac{10 z+13}{36}  \ln(z)-\frac{83}{54}
\right]
                    \nonumber \displaybreak[0] \\ &
+C_F n_f \left[
   -\frac{(1-z) \left(z^2-23 z+1\right)}{6 z}+\frac{1+z}{12}\ln^3(z)+
   \frac{3+z}{8} \ln^2(z)+\frac{3}{2} (z+1) \ln(z)
\right].
\end{align}
Here $n_f$ is the number of active quark flavors, and $\zeta_3$ is the Riemann zeta function which is defined as
\begin{equation}
    \zeta_3=\sum_{n=0}^{\infty}\frac{1}{n^3}.
\end{equation}

The above coefficients can be derived from Eq. (33) of Ref. \cite{Catani:2011kr},
\begin{eqnarray}\label{eqm}
\int_0^{Q_0^2}dq_T^2
\;\f{d{\hat \sigma}_{ab\to H}}{dq_T^2}(q_T,m_H,{\hat s})
&
\equiv& \int_0^{+\infty} dq_T^2
\;\f{d{\hat \sigma}_{ab\to H}}{dq_T^2}(q_T,m_H,{\hat s})
-\int_{Q_0^2}^{+\infty} dq_T^2
\;\f{d{\hat \sigma}_{ab\to H}}{dq_T^2}(q_T,m_H,{\hat s})
\nn \\
&=&
{\hat \sigma}_{ab\to H}^{({\rm tot})}(m_H,{\hat s})
-\int_{Q_0^2}^\infty dq_T^2 \int^{+\infty}_{-\infty} d{\hat y}
\;\f{d{\hat \sigma}_{ab\to H}}{d{\hat y} \,dq_T^2}({\hat y},q_T,m_H,{\hat s}),
\end{eqnarray}
where ${\hat \sigma}_{ab\to H}$ is the partonic cross section of Higgs boson production process $ab\to H$,
and $Q_0$ is a small cut-off scale of transverse momentum, which is taken to be much less than $m_H$.
The left hand side of Eq. (\ref{eqm}) can be expressed as
\begin{equation}
\label{inte}
\int_0^{Q_0^2}dq_T^2
\;\f{d{\hat \sigma}_{ab\to H}}{dq_T^2}(q_T,m_H,{\hat s}=m_H^2/z)
\equiv z \,\sigma_{0}
\;{\hat R}_{ab\to H}(z,m_H/Q_0) \;\;,
\end{equation}
in which $\sigma_{0}$ is the Born level partonic cross section given in Eq. (\ref{eq:sigma0}), and
\begin{align}
\label{eqr2}
{\hat R}^{(2)}_{ab\to H}(z,m_H/Q_0)&=l_0^4 \;\Sigma_{gg\ito ab}^{H(2;4)}(z)
+l_0^3\; \Sigma_{gg \ito ab}^{H(2;3)}(z)+l_0^2\; \Sigma_{gg\ito ab}^{H(2;2)}(z)
+l_0\left(\Sigma_{gg\ito ab}^{H(2;1)}(z)-16 \zeta_3 \Sigma_{gg\ito ab}^{H(2;4)}(z)\right)
\nn\\
&
+\left({\cal H}_{gg\ito ab}^{H(2)}(z)
-4\zeta_3\, \Sigma_{gg\ito ab}^{H(2;3)}(z)\right)
+{\cal O}(Q_0^2/m_H^2)\, ,
\end{align}
with $l_0=\log (m_H^2/Q_0^{2})$  \cite{Catani:2011kr}.
$\Sigma_{gg\ito ab}^{H(2;n)}, n=1,...,4$ are known in the transverse momentum resummation scheme \cite{Bozzi:2005wk}.
In the limit $Q_0\to 0$, the terms ${\cal O}(Q_0^2/m_H^2)$ can be neglected.
Moreover,
\begin{equation}
{\cal H}_{gg \ito ab}^{H (2)}\left(z \right)
=  \delta_{ga}C_{gb}^{(2)} (z)
+ \delta_{gb} C_{ga}^{(2)} (z)
+ C_{ga}^{(1)}\otimes C_{gb}^{(1)}(z)
+ G_{ga}^{(1)}\otimes G_{gb}^{(1)}(z).
\end{equation}
The calculation of ${\cal H}_{gg \ito ab}^{H (2)}(z)$ is challenging
and has been originally performed in Ref. \cite{Catani:2011kr}.
We have numerically checked Eq. (\ref{eqm}) independently using the analytical expressions
for the total cross section (the first term on the right hand side of Eq. (\ref{eqm})) in Ref. \cite{Anastasiou:2002yz}
and transverse momentum distribution (the second term on the right hand side of Eq. (\ref{eqm}))  in Ref. \cite{Glosser:2002gm},
and found some typos for  ${\cal H}_{gg \ito gq}^{H (2)}$ in the first version of Ref. \cite{Catani:2011kr}
\footnote{ After our paper appeared on the arXiv, the expression of  ${\cal H}_{gg \ito gq}^{H (2)}$
in Ref. \cite{Catani:2011kr} was corrected in its revised version. },
while the expression of  ${\cal H}_{gg \ito gq}^{H (2)}$ in the program HNNLO \cite{Catani:2007vq,Grazzini:2008tf} is correct.

The function $\widetilde{W}_{gg}^{NP}$ describes the non-perturbative part of the soft gluon resummation.
We use the BLNY parameterization form \cite{Landry:2002ix},
with the non-perturbative coefficients scaled by the factor of $C_A/C_F=9/4$.
This scaling factor is applied since the initial states are gluons in the $gg \rightarrow H$ process,
instead of quarks in the Drell-Yan process.
The effect of varying the non-perturbative coefficients will be discussed in the next section.

In addition, the G-function, which is only present for the gluon initiated processes,
can be  incorporated into the CSS formula as shown in Ref. \cite{Catani:2011kr}.
Therefore, the term $\widetilde W$ can be further improved as
\begin{equation}\label{G-function0}
\widetilde{W}_{gg}(b,Q,x_{1},x_{2},C_{1,2,3}) =
e^{-S(b,Q,C_1,C_2)}\sum_{a,b=q,\bar{q},g}\left[
\bigl(C_{ga}\otimes f_a\bigr)(x_1)\bigl(C_{gb}\otimes f_b\bigr)(x_2)
+\bigl(G_{ga}\otimes f_a\bigr)(x_1)\bigl(G_{gb}\otimes f_b\bigr)(x_2)
\right]
.
\end{equation}
The lowest order of G-function is at ${\mathcal O}(\alpha_s/\pi)$, and
\begin{equation}\label{G-function}
G_{ga}(z,\alpha_s)=\frac{\alpha_s}{\pi}G_{ga}^{(1)}(z)+\sum_{n=2}^\infty\left(\frac{\alpha_s}{\pi}\right)^n
G_{ga}^{(n)}(z),
\end{equation}
where the coefficient functions are known as \cite{Nadolsky:2007ba,Catani:2011kr,Catani:2010pd}
\begin{equation}\label{G-function1}
G_{gg}^{(1)}(z)=C_A\frac{1-z}{z}, \quad G_{gq}^{(1)}(z)=C_F\frac{1-z}{z},
\end{equation}
in which $z$ denotes the fraction of momentum  carried away by the gluon.

Finally, in Eq. (\ref{eq:resum_formalism}), the term containing $Y$ incorporates
the remainder of the cross section which is not singular as $Q_T\rightarrow 0$.
It consists of the difference between the full fixed-order cross section  and the small-$Q_T$ limit of this cross section.
At small $Q_T$, they cancel with each other, so that the contribution of the $Y$-term is small,
and the $\widetilde{W}$ term dominates.
At large $Q_T$, where the logarithmic terms become small,
the $\widetilde{W}$ term cancels against the small-$Q_T$ limit term (to the given order in $\alpha_s$),
so that the result approaches the fixed-order calculations.
More details of how this matching process between the resummation calculation and the fixed-order one
is implemented in the ResBos program can be found in Ref. \cite{Balazs:1997xd}.

\section{Numerical Results}
\label{sec:nume}

In this section, we investigate the effect of including the NNLO Wilson coefficient functions and G-function contributions
in the ResBos2 calculations on the  total cross sections and transverse momentum distributions.
Unless otherwise specified, the parton distribution function (PDF) used in the numerical calculations
is chosen as CTEQ6.6 \cite{Nadolsky:2008zw}.
The scale uncertainty is obtained  by varying the scale $C_2$ around the canonical choice ($C_2=1$)
by a factor of two, but maintaining the relations
\begin{equation}\label{canonical}
 C_1=C_2 b_0, \quad C_3=b_0 \quad {\rm and} \quad  C_4=C_2.
 \end{equation}
Keeping these relations between $C_i, i=1,2,3,4$, the Wilson coefficient functions are not changed.
Therefore, the dominant effect of the variation is to change the shape,
rather than the overall size of the cross section.
We shall compare the  ResBos2 predictions  with those from the programs HNNLO
and HqT2 on the total cross sections and the transverse momentum distributions.

\begin{table}[!htb]
\begin{tabular}{c|c|c|c|c|c}
\hline
& $m_H$ (GeV) & ResBos2 & ResBos & HNNLO (NNLO) & HqT2 (NNLL+NLO)
\\ \hline
\multirow{4}{*}{Tevatron}
 & 115 &  $0.98^{+9.2\%}_{-6.1\%}$   & $0.91^{+15.7\%}_{-6.9\%}$ & $0.96^{+13.6\%}_{-13.7\%}$ & $0.97^{+14.2\%}_{-13.8\%}$
\tabularnewline \cline{2-6}
 & 120 &  $0.87^{+9.2\%}_{-6.9\%}$   & $0.80^{+15.8\%}_{-6.6\%}$ & $0.84^{+13.4\%}_{-13.7\%}$ & $0.85^{+13.9\%}_{-13.8\%}$
\tabularnewline \cline{2-6}
 & 125 &  $0.77^{+9.1\%}_{-6.5\%}$   & $0.71^{+15.9\%}_{-6.5\%}$ & $0.75^{+13.5\%}_{-14.2\%}$ & $0.75^{+13.8\%}_{-13.8\%}$
\tabularnewline \cline{2-6}
 & 130 &  $0.68^{+10.3\%}_{-5.9\%}$  & $0.63^{+15.9\%}_{-6.4\%}$ & $0.66^{+14.5\%}_{-13.1\%}$ & $0.67^{+13.9\%}_{-13.8\%}$
\\ \hline
\multirow{4}{*}{LHC 7 TeV}
 & 115 &  $15.11^{+7.8\%}_{-5.8\%}$  & $14.21^{+8.3\%}_{-5.9\%}$ & $15.16^{+9.0\%}_{-10.7\%}$  & $15.19^{+10.8\%}_{-10.4\%}$
\tabularnewline \cline{2-6}
 & 120 &  $13.89^{+7.8\%}_{-5.7\%}$  & $13.06^{+8.4\%}_{-5.8\%}$ & $13.80^{+10.6\%}_{-10.5\%}$ & $13.94^{+10.2\%}_{-10.4\%}$
\tabularnewline \cline{2-6}
 & 125 &  $12.80^{+7.7\%}_{-5.5\%}$  & $12.03^{+8.5\%}_{-5.6\%}$ & $12.72^{+10.2\%}_{-10.6\%}$ & $12.83^{+10.7\%}_{-10.4\%}$
\tabularnewline \cline{2-6}
 & 130 &  $11.83^{+7.7\%}_{-5.4\%}$  & $11.12^{+8.6\%}_{-5.5\%}$ & $11.75^{+10.8\%}_{-10.7\%}$ & $11.84^{+9.9\%}_{-10.4\%}$
\\ \hline
\multirow{4}{*}{LHC 8 TeV}
 & 115 &  $19.15^{+7.5\%}_{-6.5\%}$  & $17.58^{+8.1\%}_{-7.1\%}$ & $19.05^{+9.9\%}_{-9.2\%}$    & $19.25^{+10.8\%}_{-9.8\%}$
\tabularnewline \cline{2-6}
 & 120 &  $17.65^{+7.5\%}_{-6.5\%}$  & $16.21^{+8.1\%}_{-7.1\%}$ & $17.59^{+9.7\%}_{-9.9\%}$    & $17.76^{+9.8\%}_{-10.1\%}$
\tabularnewline \cline{2-6}
 & 125 &  $16.31^{+7.5\%}_{-6.4\%}$  & $14.98^{+8.1\%}_{-7.0\%}$ & $16.26^{+10.2\%}_{-10.4\%}$  & $16.39^{+9.8\%}_{-10.1\%}$
\tabularnewline \cline{2-6}
 & 130 &  $15.11^{+7.5\%}_{-6.4\%}$  & $13.89^{+8.1\%}_{-7.0\%}$ & $15.07^{+10.9\%}_{-10.6\%}$  & $15.17^{+10.3\%}_{-10.1\%}$
\\ \hline
\multirow{4}{*}{LHC 14 TeV}
& 115 &  $48.84^{+7.6\%}_{-5.7\%}$  & $46.03^{+10.0\%}_{-5.3\%}$ & $49.24^{+9.4\%}_{-9.8\%}$ & $48.90^{+9.0\%}_{-9.8\%}$
\tabularnewline \cline{2-6}
& 120 &  $45.45^{+7.5\%}_{-5.5\%}$  & $42.84^{+9.8\%}_{-5.2\%}$ & $45.57^{+10.4\%}_{-9.5\%}$ & $45.52^{+10.3\%}_{-8.4\%}$
\tabularnewline \cline{2-6}
& 125 &  $42.39^{+7.4\%}_{-5.4\%}$  & $39.96^{+9.6\%}_{-5.0\%}$ & $42.61^{+9.6\%}_{-9.7\%}$  & $42.57^{+9.7\%}_{-8.8\%}$
\tabularnewline \cline{2-6}
& 130 &  $39.65^{+7.3\%}_{-5.2\%}$  & $37.38^{+9.4\%}_{-4.8\%}$ & $39.93^{+11.1\%}_{-9.8\%}$ & $39.75^{+9.7\%}_{-8.9\%}$
\\ \hline
\end{tabular}
\caption{The ResBos2, ResBos, HNNLO and HqT2 predictions on the total cross
sections (in pb) for Higgs boson production via $g+g\to H+X$ at the Tevatron (1.96 TeV) and LHC  (7 TeV, 8 TeV and 14 TeV).
The upper (lower) uncertainties, expressed in the form of percentages, are obtained by dividing (multiplying) the canonical scale by a factor of two.}
\label{TOTXsec}
\end{table}

Table \ref{TOTXsec} shows the  predictions on the total cross sections for Higgs boson production at the Tevatron and the LHC.
By including the NNLO Wilson coefficient function $C^{(2)}$,
ResBos2 predictions at the Tevatron and the 7 TeV (8 TeV, 14 TeV) LHC are larger than ResBos predictions by about
$8\%$ and $6\%$ ($9\%$, $6\%$), respectively.
The theoretical uncertainties of ResBos (ResBos2) predictions are estimated by varying the scales in the resummation formula,
as discussed around Eq. (\ref{canonical}).
They are reduced in ResBos2 to about $9\%$ and $8\%$ ($7\%$, $8\%$) at the Tevatron and the 7 TeV (8 TeV, 14 TeV) LHC, respectively.

We can also see that ResBos2 predictions agree with the predictions from HNNLO and  HqT2 programs within a couple of percent.
When the CM energy of the LHC is increased from 7 TeV to 8 TeV,
the total cross sections for the Higgs boson production are increased by about $27\%$ for $m_H$=125 GeV.

\begin{table}[!htb]
\begin{tabular}{c|c|c|c|c|c|c}
\hline
&  CTEQ6.6 & CT10 NLO & CT10W NLO & CT10 NNLO & MSTW2008NNLO & NNPDF2.3NNLO
\\ \hline
 Tevatron  &$0.77\pm 7.2\%$ &  $0.77\pm 7.2\%$   & $0.76\pm 7.1\%$ & $0.77\pm 7.2\%$
& $0.78\pm6.6\%$ & $0.80\pm4.5\%$
\\ \hline
LHC 7 TeV  & $12.80\pm 3.3\%$ & $13.33\pm 3.3\%$  & $12.82\pm 3.3\%$  & $12.65\pm 2.9\%$
& $12.69\pm 2.7\%$ & $13.73\pm 2.3\%$
\\ \hline
 LHC 8 TeV  & $16.31\pm 3.4\%$ & $16.53\pm 3.4\%$  & $16.95\pm 3.4\%$ & $16.63\pm 3.1\%$
& $16.30\pm 2.4\%$ & $16.90\pm 2.1\%$
\\ \hline
LHC 14 TeV  & $42.39\pm 3.5\%$ & $42.64\pm 3.5\%$  & $42.91\pm 3.4\%$  & $41.87\pm 3.6\%$
& $43.10\pm 2.3\%$ & $43.28\pm 2.0\%$
\\ \hline
\end{tabular}
\caption{The total cross sections (in pb) for Higgs boson production via $g+g\to H+X$ at the Tevatron (1.96 TeV) and LHC  (7 TeV, 8 TeV and 14 TeV)
by using different PDF sets in ResBos2.
The PDF induced uncertainties are estimated at (or rescaled to) 90$\%$ confidence-level, and expressed in the form of percentages.}
\label{PDFerror}
\end{table}

To study the dependence of ResBos2 predictions on the choice of PDF sets,
 we show in Table \ref{PDFerror} the total cross sections of producing a 125 GeV Higgs boson  at the Tevatron and the LHC,
and corresponding PDF uncertainties when using the CTEQ6.6, CT10 NLO, CT10W NLO, CT10 NNLO \cite{Lai:2010vv,Nadolsky:2012ia},
MSTW2008NNLO \cite{Martin:2009iq} and NNPDF2.3NNLO \cite{Nocera:2012hx} PDF sets.
We compare these widely used PDF sets as recommended by the PDF4LHC Working Group in Ref. \cite{Botje:2011sn}.
There is an about $8\%$ difference between the NNPDF2.3NNLO and CT10 NNLO predictions at the 7 TeV LHC.
In general, the NNPDF2.3NNLO sets predict larger cross sections for the Higgs boson production than the other PDF sets.
A similar conclusion is also found in Ref. \cite{Baglio:2012et}.

\begin{figure}[!htb]
\includegraphics[width=0.45\textwidth]{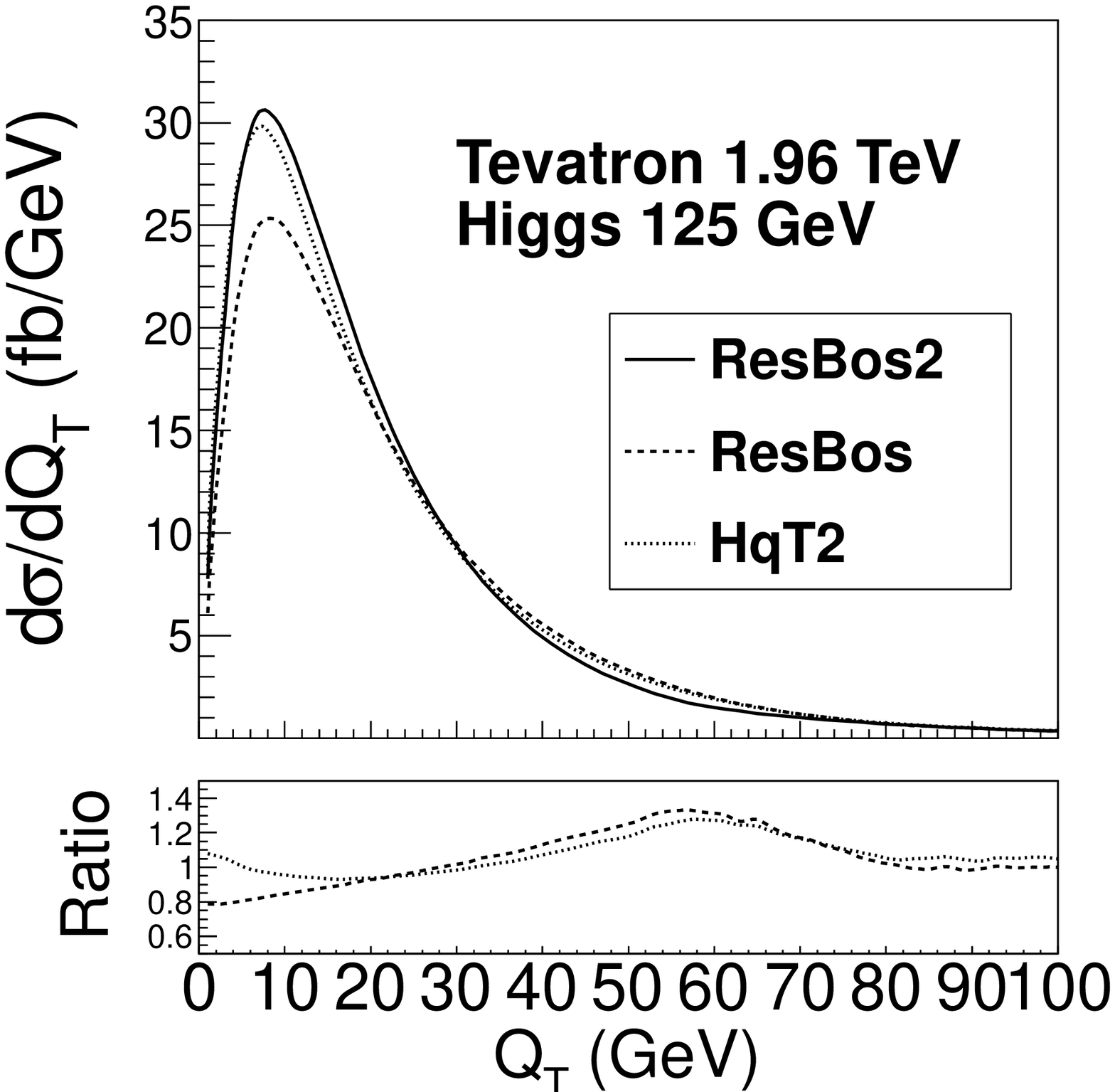}
\includegraphics[width=0.45\textwidth]{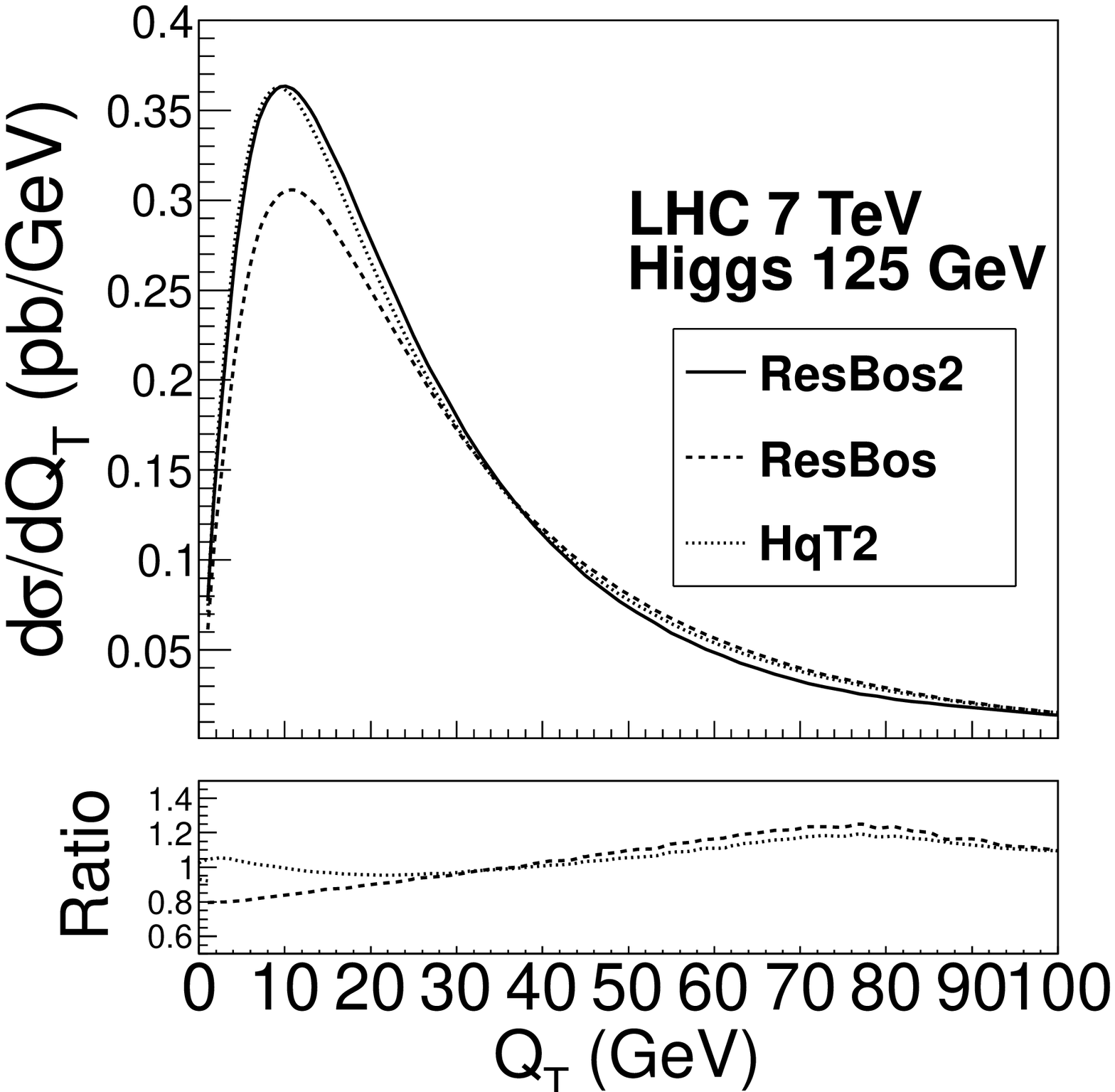}
\includegraphics[width=0.45\textwidth]{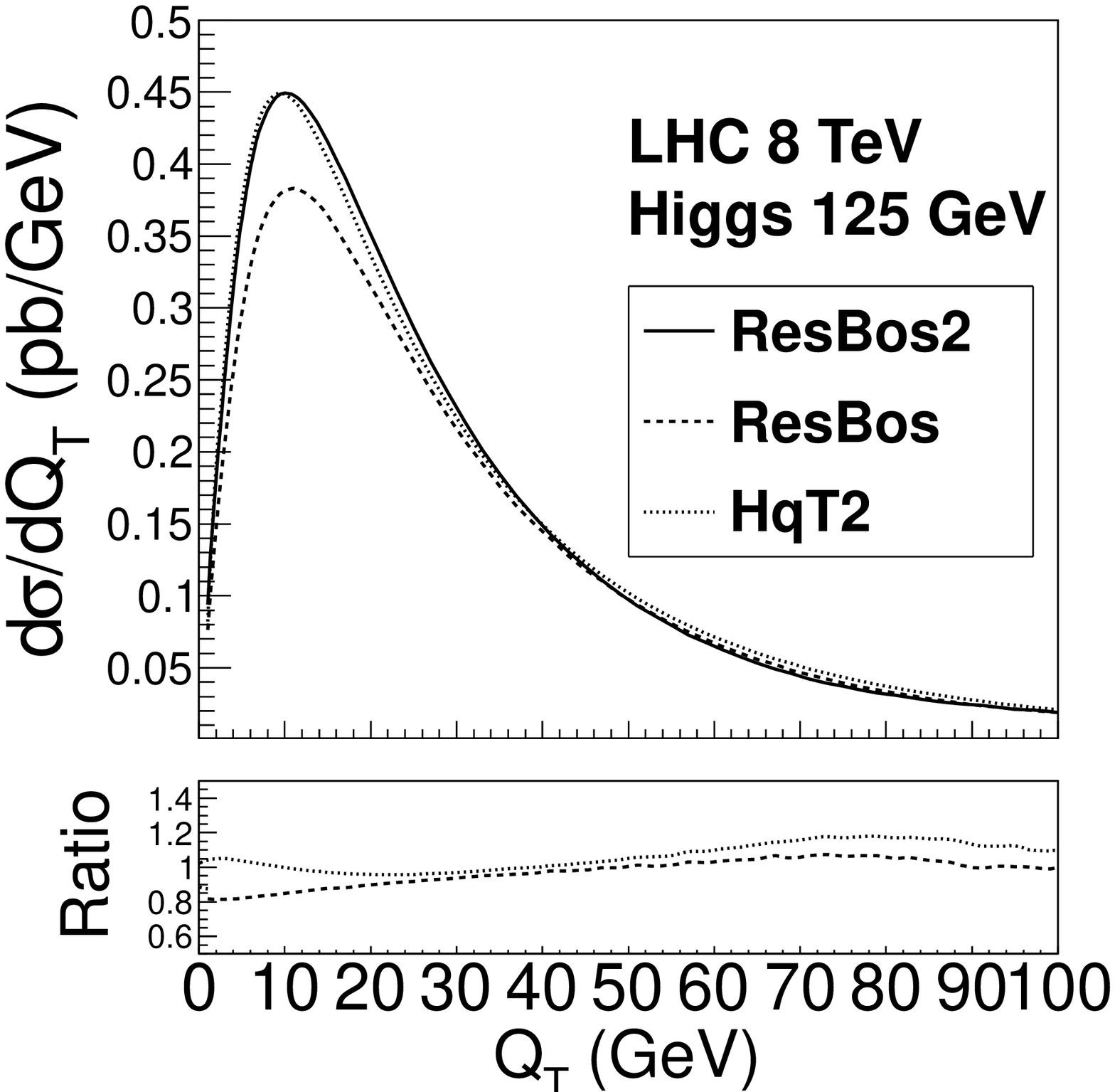}
\includegraphics[width=0.45\textwidth]{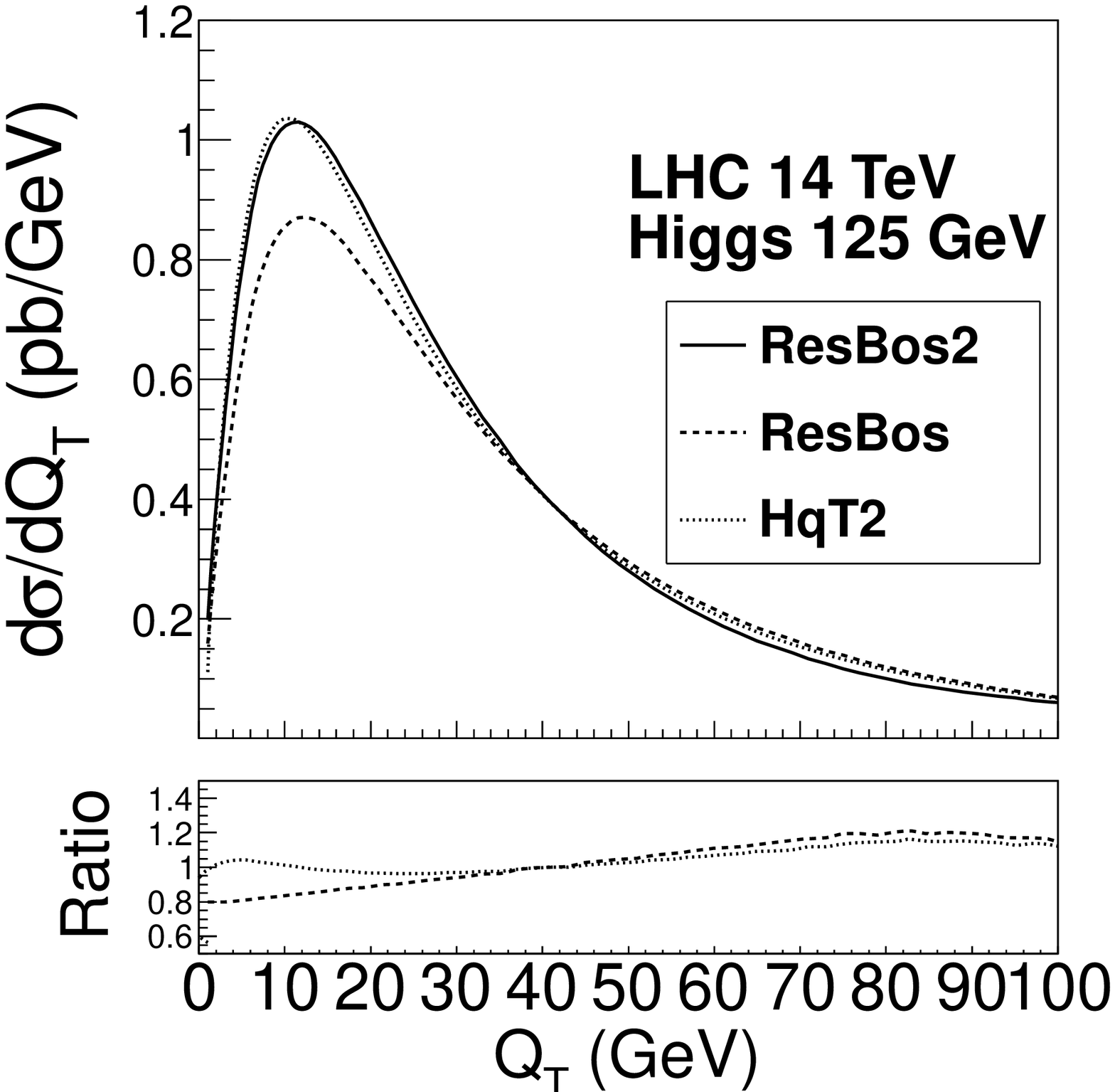}
\caption{The different theoretical predictions on the
transverse momentum distributions for the Higgs boson production at the Tevatron (1.96 TeV)
and the LHC (7 TeV, 8 TeV and 14 TeV).
In the bottom of each plot, the ratios to ResBos2 predictions are also shown.}
\label{ptdistCOMPARE}
\end{figure}

In Fig. \ref{ptdistCOMPARE}, we present ResBos2 predictions on the transverse momentum distributions of the Higgs boson,
compared with the predictions of ResBos and HqT2.
The predictions of ResBos2 generally increase the overall size of the distributions, compared to ResBos.
Especially, the improvement in the small transverse momentum region can be as large as $20\%$ at both the Tevatron and the LHC.
The peak positions are slightly shifted to small $Q_T$ region.
Meanwhile, ResBos2 predictions are smaller than ResBos by about $20\%$ for $Q_T\sim 60$ GeV ($80$ GeV) at the Tevatron (LHC).
This behavior is due to the fact that the inclusion of NNLO Wilson coefficient functions
modifies the resummed ($\tilde W$) contribution and hence the matching condition in the intermediate $Q_T$ region.
We note that the high $Q_T$ predictions remain to be the same.
The HqT2 predictions at the LHC (7 TeV, 8 TeV and 14 TeV) in small $Q_T$ region are almost the same as ResBos2,
but the peak height of the HqT2 prediction at the Tevatron is a little lower than that of ResBos2.
On the other hand, the HqT2 predictions at moderate $Q_T$ region are closer to the ResBos predictions,
and can be about $20\%$ higher than ResBos2 predictions for $Q_T\sim 60$ GeV ($80$ GeV) at the Tevatron (LHC).

\begin{figure}[!htb]
\includegraphics[width=0.45\textwidth]{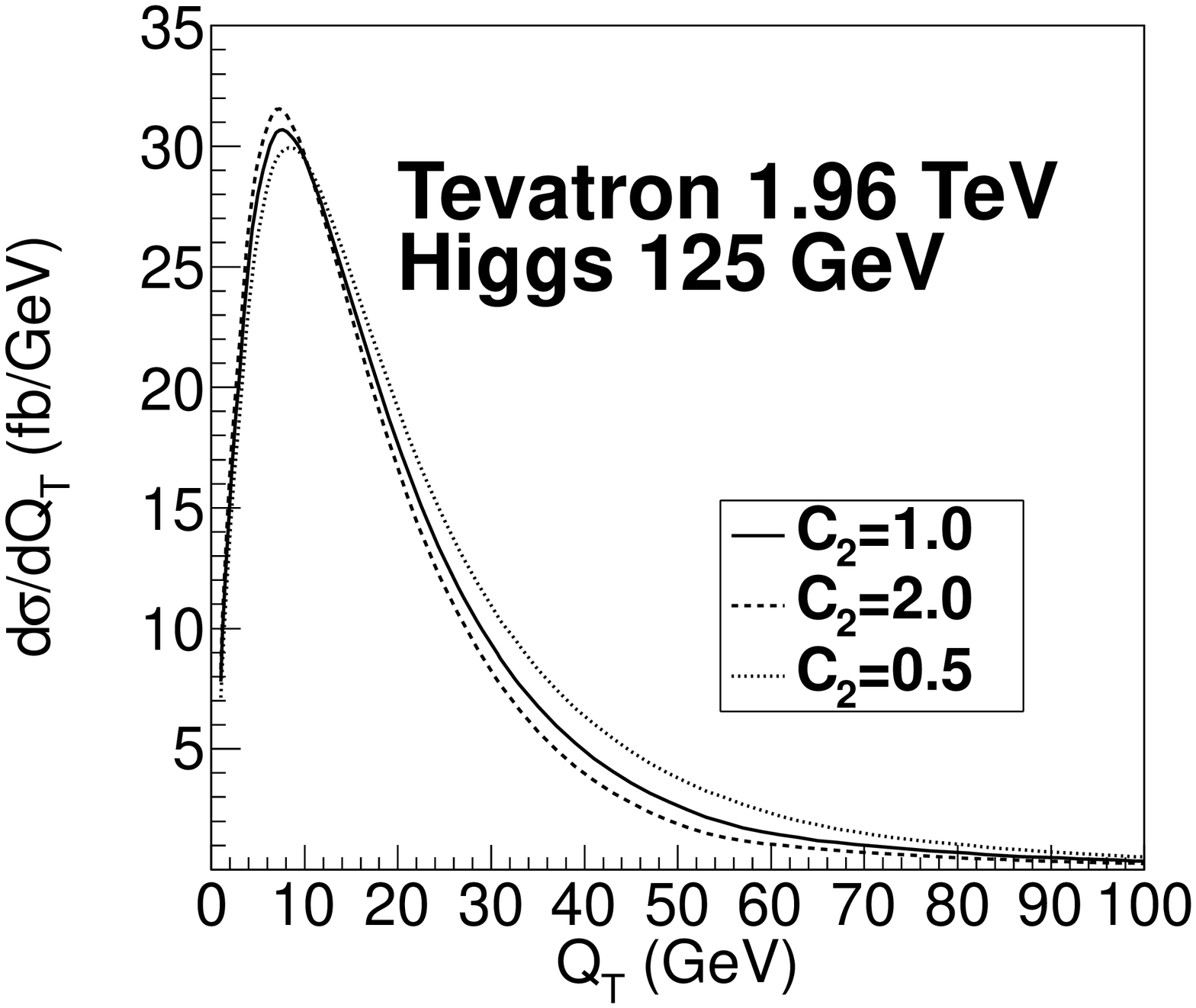}
\includegraphics[width=0.45\textwidth]{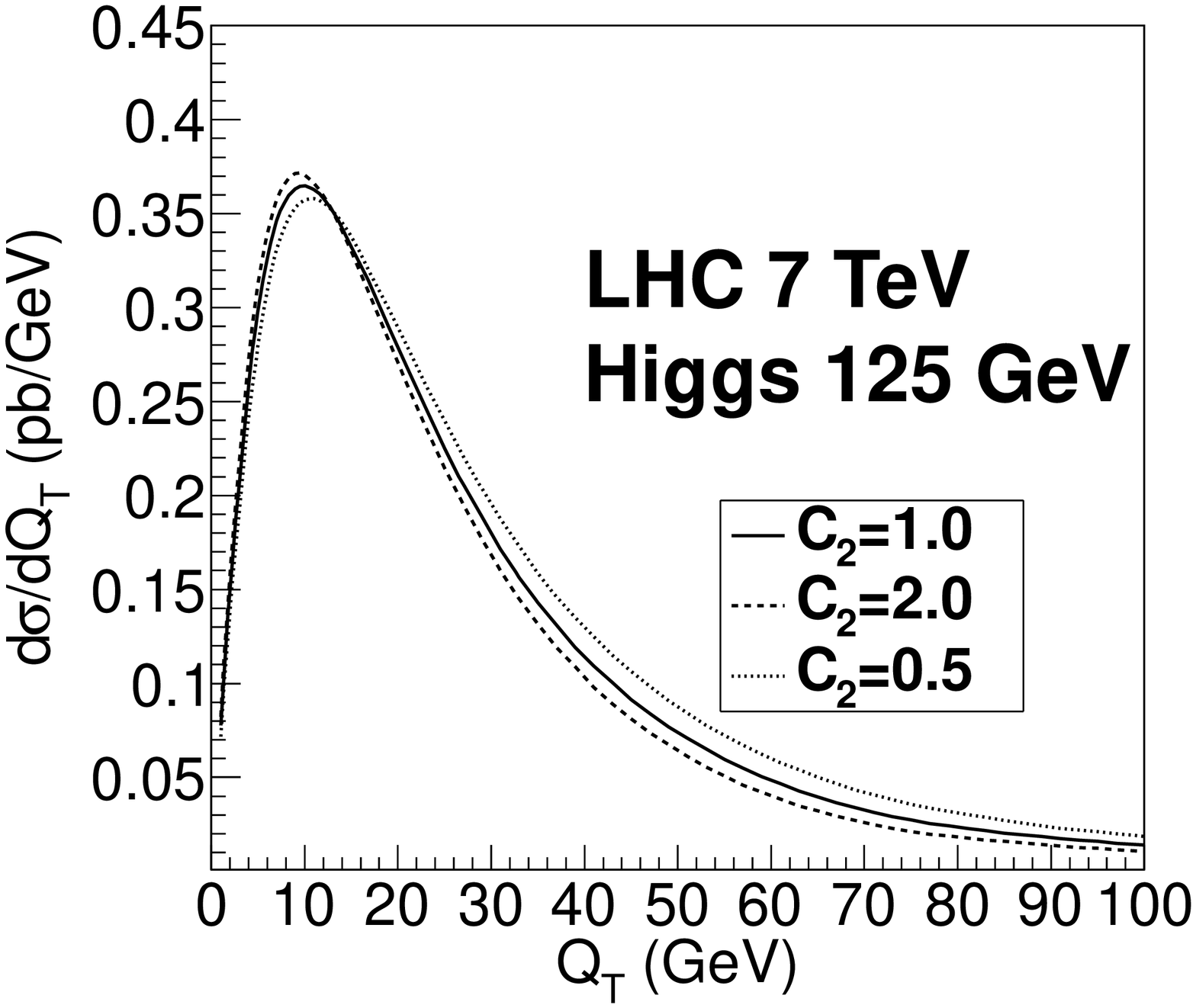}
\includegraphics[width=0.45\textwidth]{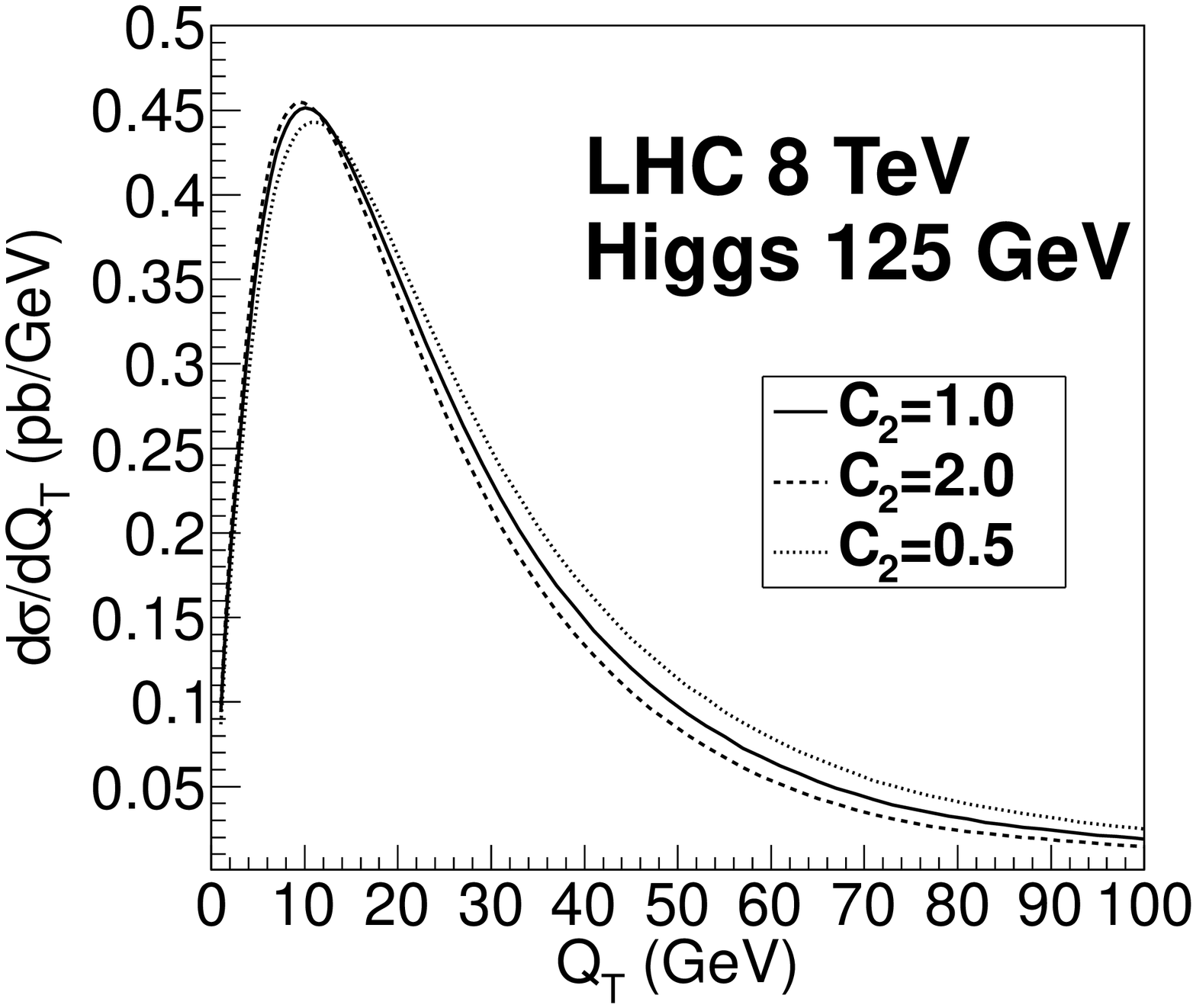}
\includegraphics[width=0.45\textwidth]{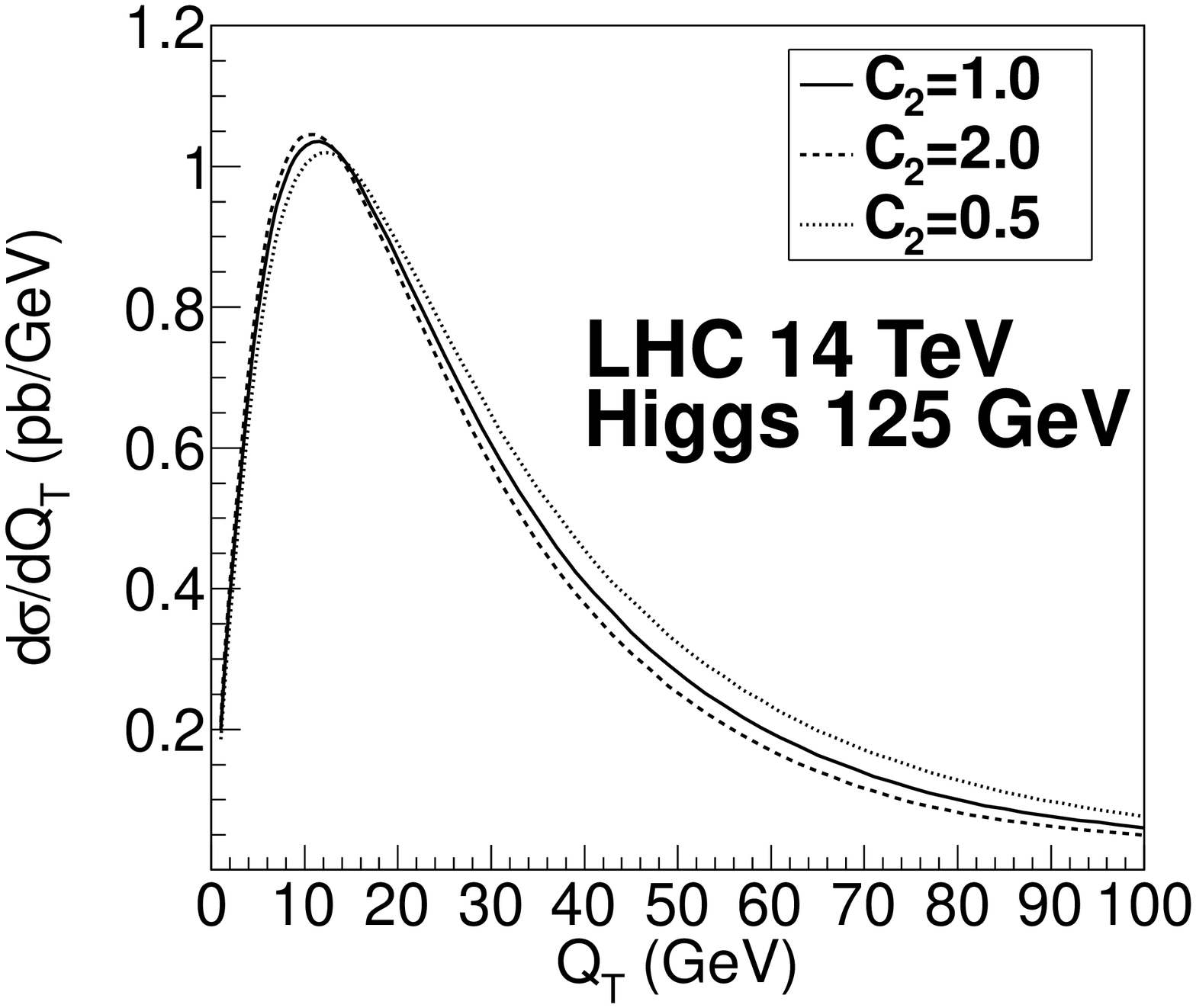}
\caption{The transverse momentum distributions of the Higgs boson for different scale choices at the Tevatron (1.96 TeV)
and the LHC (7 TeV, 8 TeV and 14 TeV).}
\label{ptdistRESBOS}
\end{figure}

Figure \ref{ptdistRESBOS} shows the resummation scale dependencies of the transverse momentum distributions
of the Higgs boson produced  at the Tevatron and the LHC.
To estimate the resummation scale dependencies, we vary the hard scale coefficient $C_2$ by a factor of two
around the canonical choice, but hold the relations between $C_{i},i=1,2,3,4$, as shown in Eq. (\ref{canonical}).
With this choice, the Wilson coefficient functions $C_{gg}$ and $C_{gq}$ do not change when $C_2$ varies.
Therefore, mainly the shapes of distributions are affected by varying the resummation scales.
As shown in Fig. \ref{ptdistRESBOS}, the shapes of the transverse momentum distributions of
Higgs boson are changed at both the Tevatron and the LHC.
The peak positions of the distributions are shifted by several GeV's when the scales are varied.
Generally, the distributions in the low and high $Q_T$ regions are increased  and suppressed, respectively,
when  $C_2$ increases.

\begin{figure}[!htb]
\includegraphics[width=0.45\textwidth]{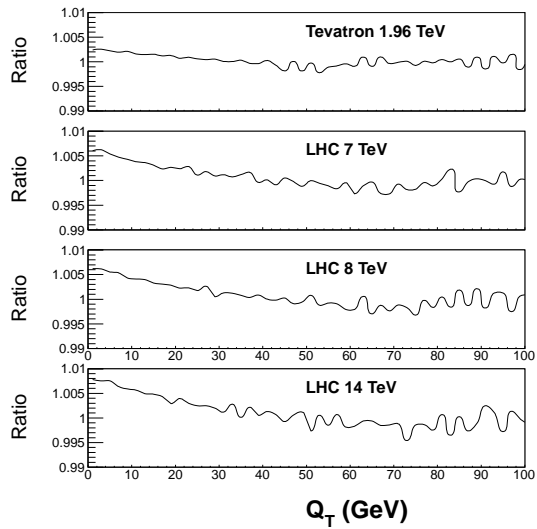}
\caption{The ratios between the transverse momentum distributions with and without G-functions
at the Tevatron (1.96 TeV) and the LHC (7 TeV, 8 TeV and 14 TeV).
The oscillations of the ratio curves in the figure are due to numerical uncertainties.}
\label{RatioRESBOS}
\end{figure}

Figure \ref{RatioRESBOS} shows the ratios between the transverse momentum distributions with and without
G-functions at the Tevatron and the LHC.
It can be seen that the effect of G-functions is very small.
As shown in the figure, by including G-functions,
the shapes of transverse momentum distributions are slightly changed, by less than $1\%$.
This small effect is understood, for the contributions of G-functions to the transverse momentum distribution
of the Higgs boson start at $\mathcal{O}(\alpha_s^4)$ [see Eqs. (\ref{G-function0})-(\ref{G-function1})].
Note that the LO cross section is at $\mathcal{O}(\alpha_s^2)$.

\begin{figure}[!htb]
\includegraphics[width=0.45\textwidth]{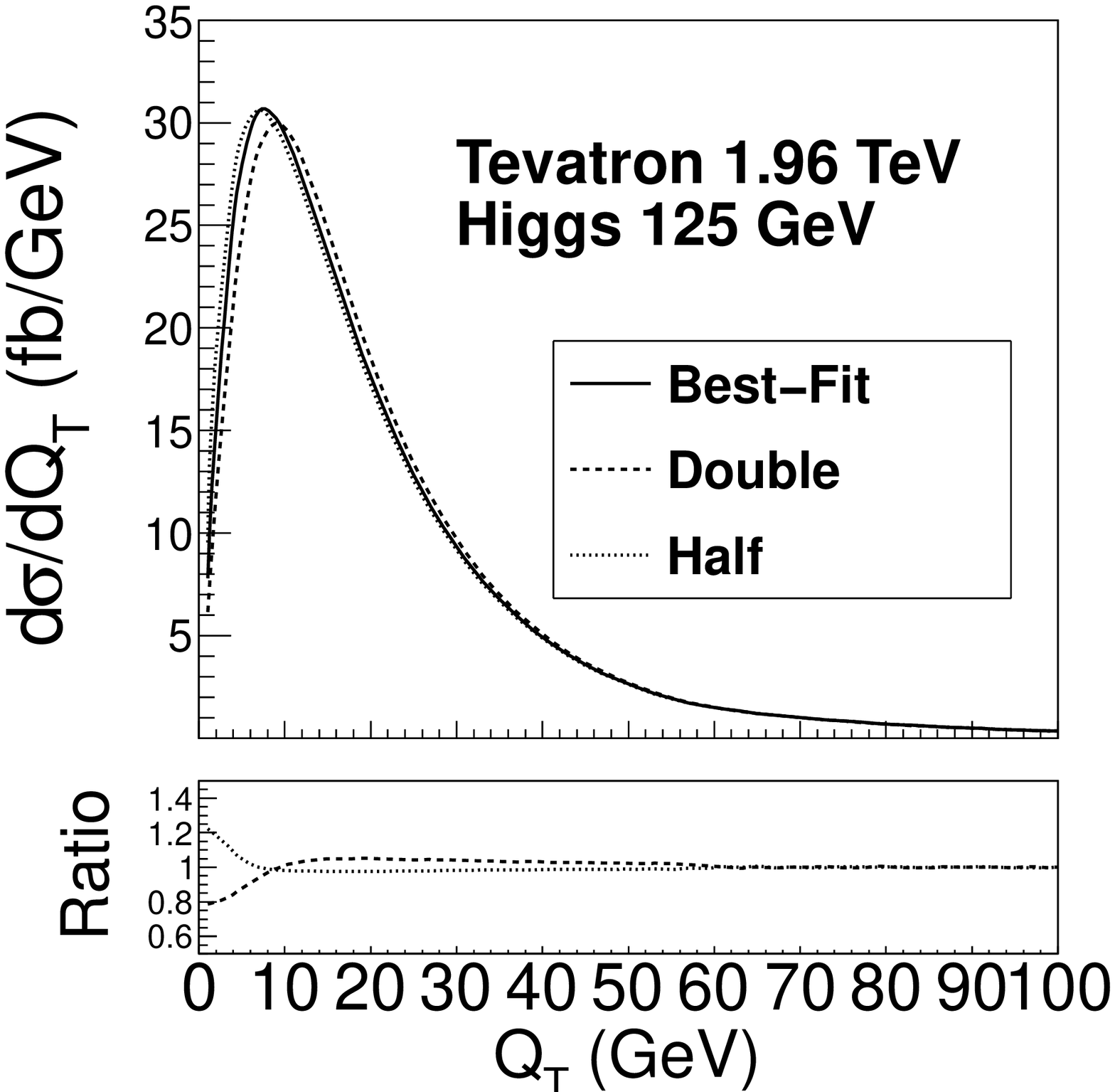}
\includegraphics[width=0.45\textwidth]{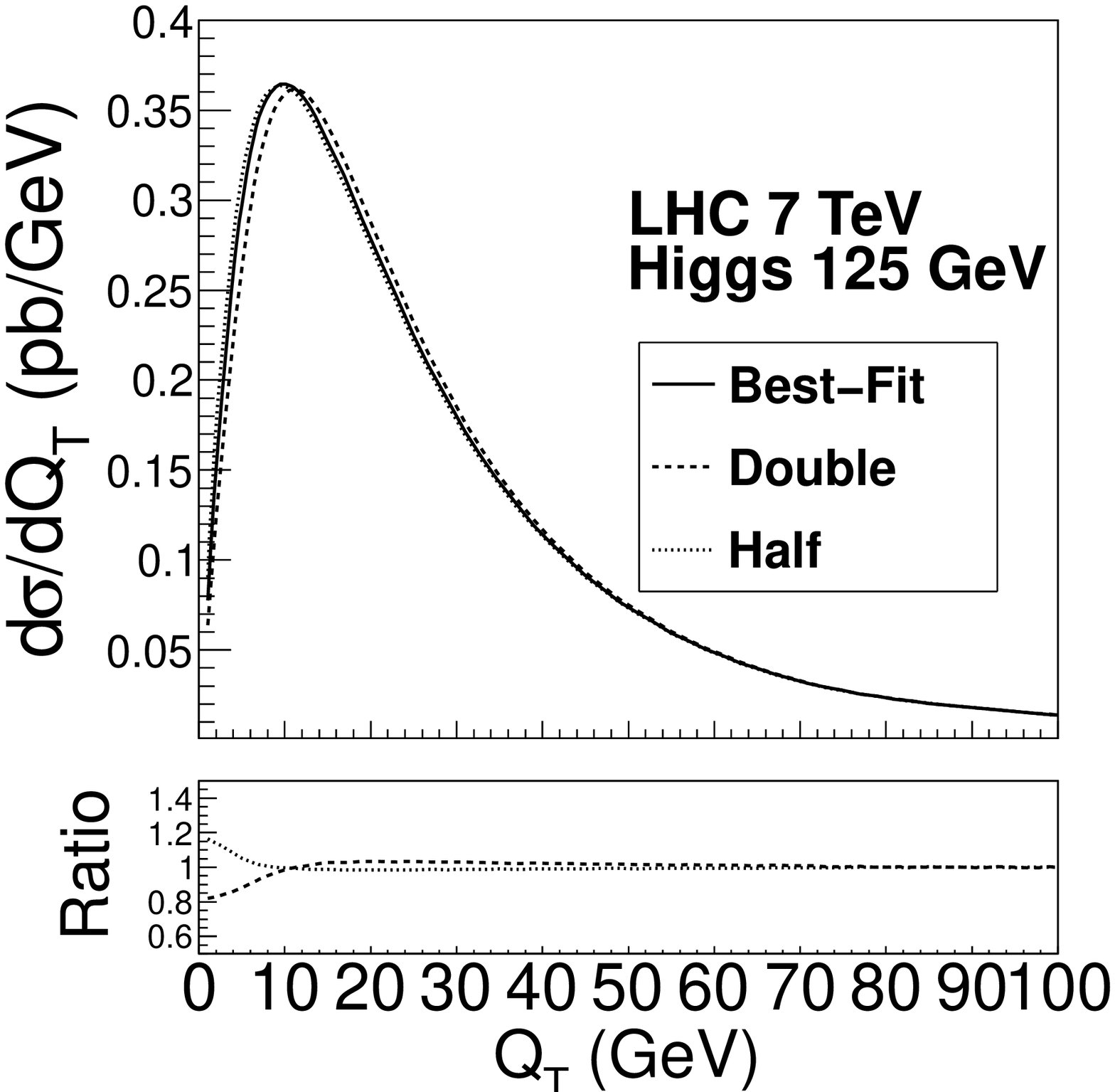}
\includegraphics[width=0.45\textwidth]{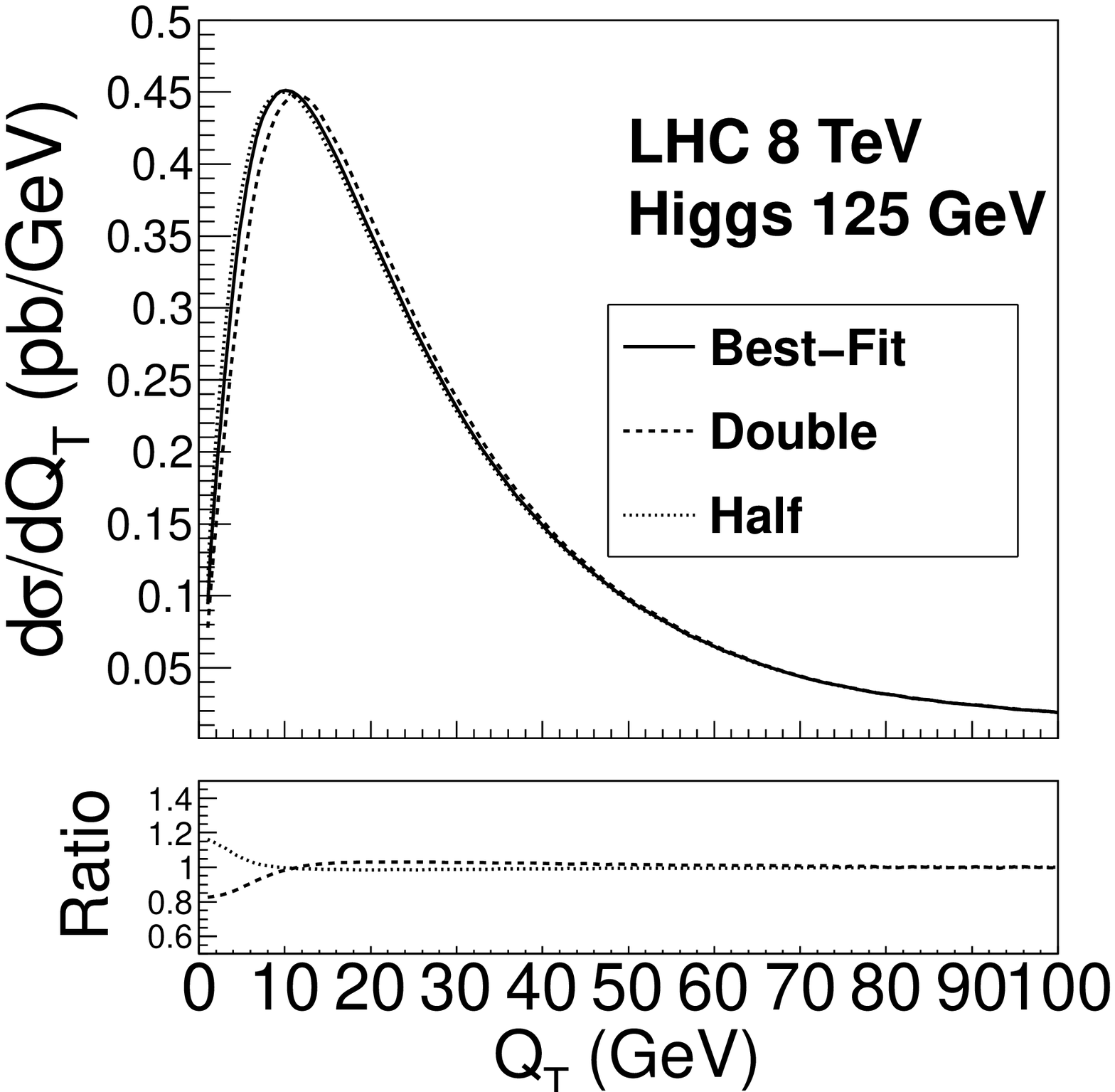}
\includegraphics[width=0.45\textwidth]{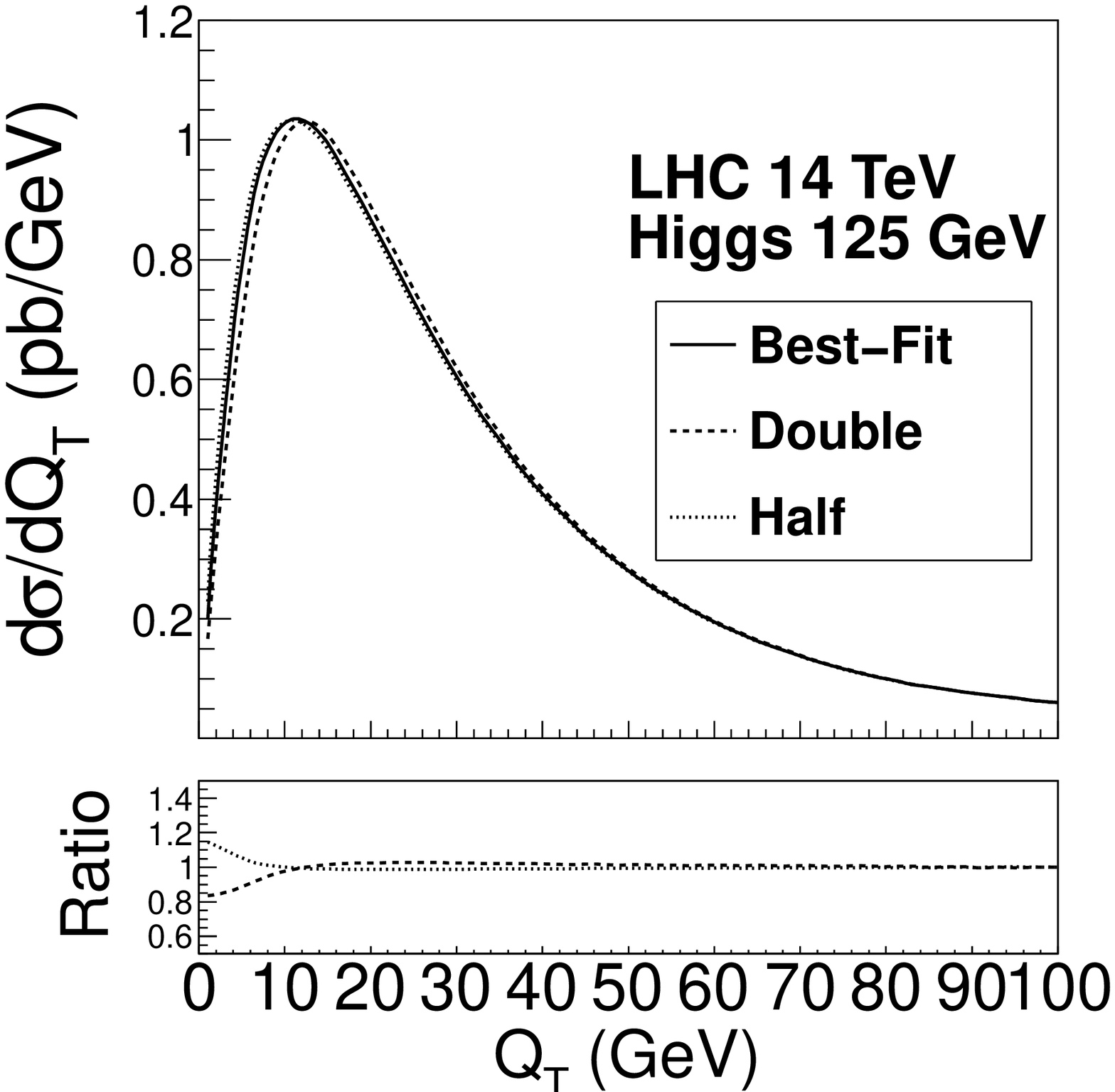}
\caption{The transverse momentum distributions for different non-perturbative parameter choices at the Tevatron (1.96 TeV)
and the LHC (7 TeV, 8 TeV and 14 TeV).}
\label{ptdistNP}
\end{figure}

Figure \ref{ptdistNP} presents the dependencies of  the transverse momentum distributions
on non-perturbative coefficients in soft gluon resummation at the Tevatron and the LHC.
Based on the best-fit non-perturbative coefficients in the BLNY parameterization form \cite{Landry:2002ix},
but scaled by the ratio $C_A/C_F=9/4$,
the dependencies on the non-perturbative coefficients are shown by multiplying or dividing
the best-fit values by a factor of two.
We can see that when the values of the non-perturbative coefficients increase,
the peak positions are slightly shifted towards larger $Q_T$ at both the Tevatron and the LHC,
and the peak heights are decreased by a few percent at the Tevatron but hardly change at the  LHC.
On the other hand, when the values of the non-perturbative coefficients decrease,
the peak positions slightly move towards smaller $Q_T$,
but the peak heights remain the same at both the Tevatron and the LHC.
Hence, we conclude that the transverse momentum distributions of the Higgs boson produced at the LHC
are dominated by the perturbative Sudakov resummation effect of multiple soft gluon emissions,
and the effects of non-perturbative physics are not as important.

In the very low $Q_T$ region, the non-perturbative physics becomes dominant,
whose effect in the prediction of the Higgs boson transverse momentum distributions can be better estimated
after the non-perturbative parameters (and the function form) are measured
from the gluon initiated scattering processes, such as top quark pair production at the LHC.

\begin{figure}[!htb]
\includegraphics[width=0.45\textwidth]{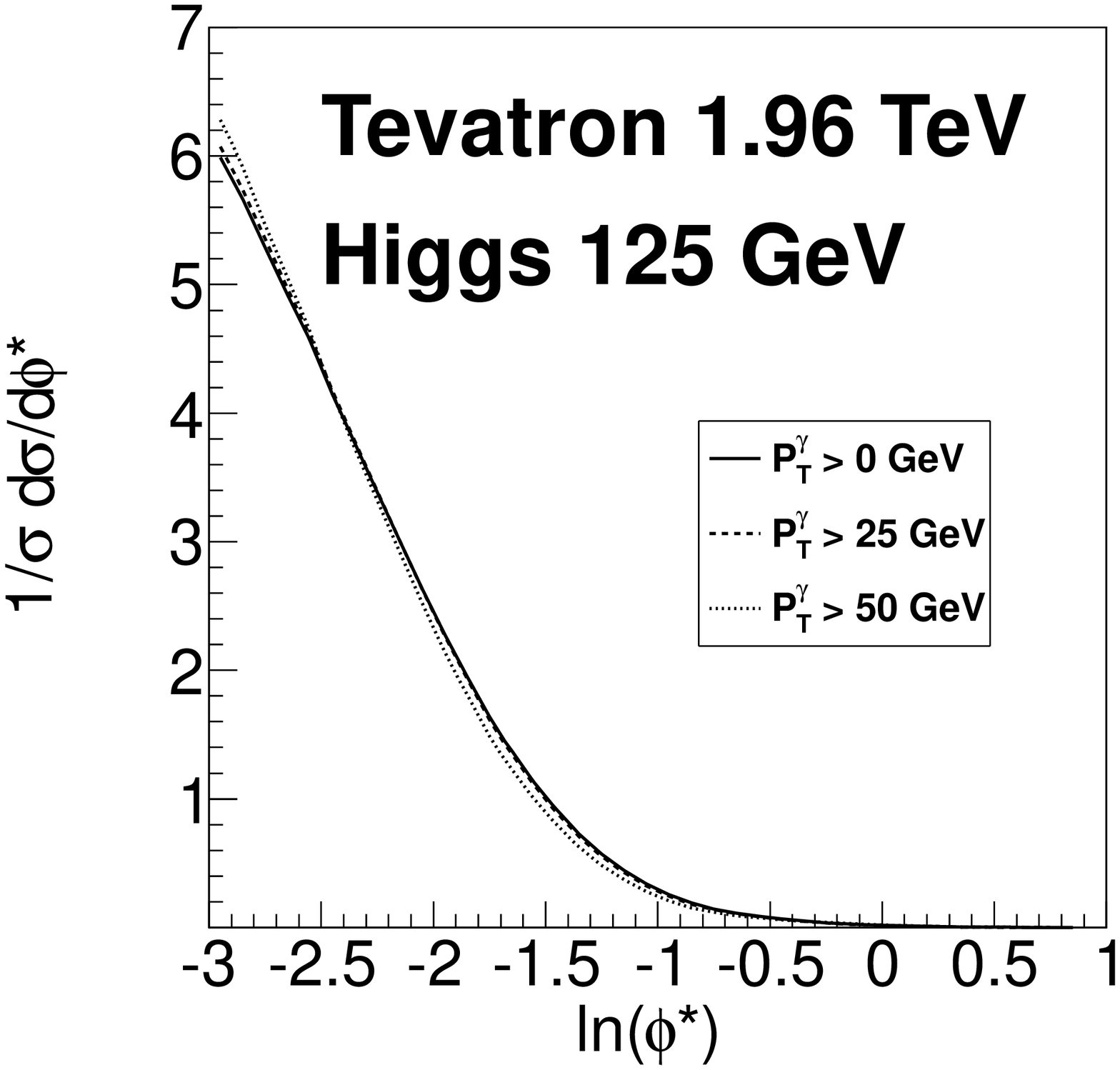}
\includegraphics[width=0.45\textwidth]{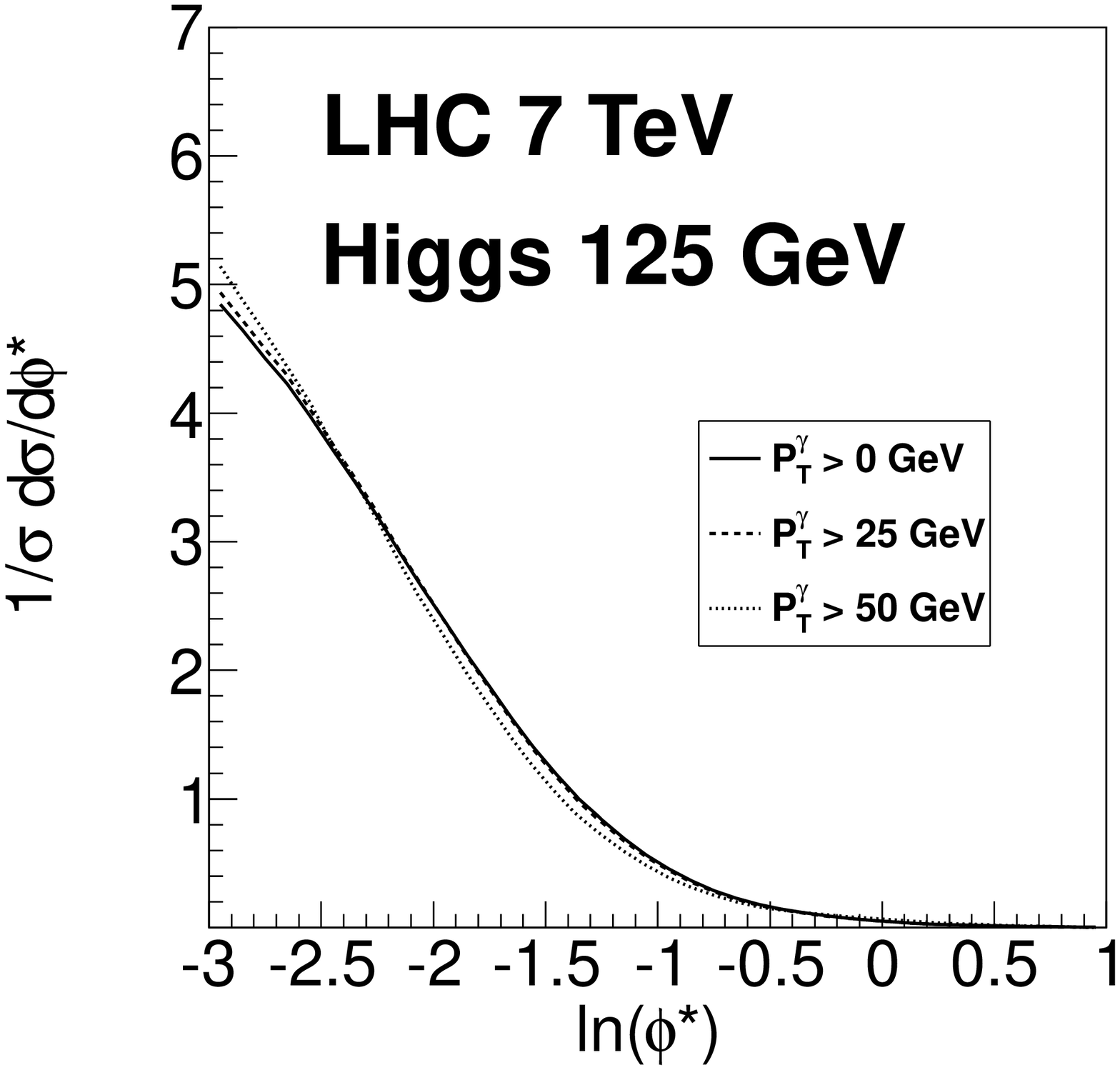}
\includegraphics[width=0.45\textwidth]{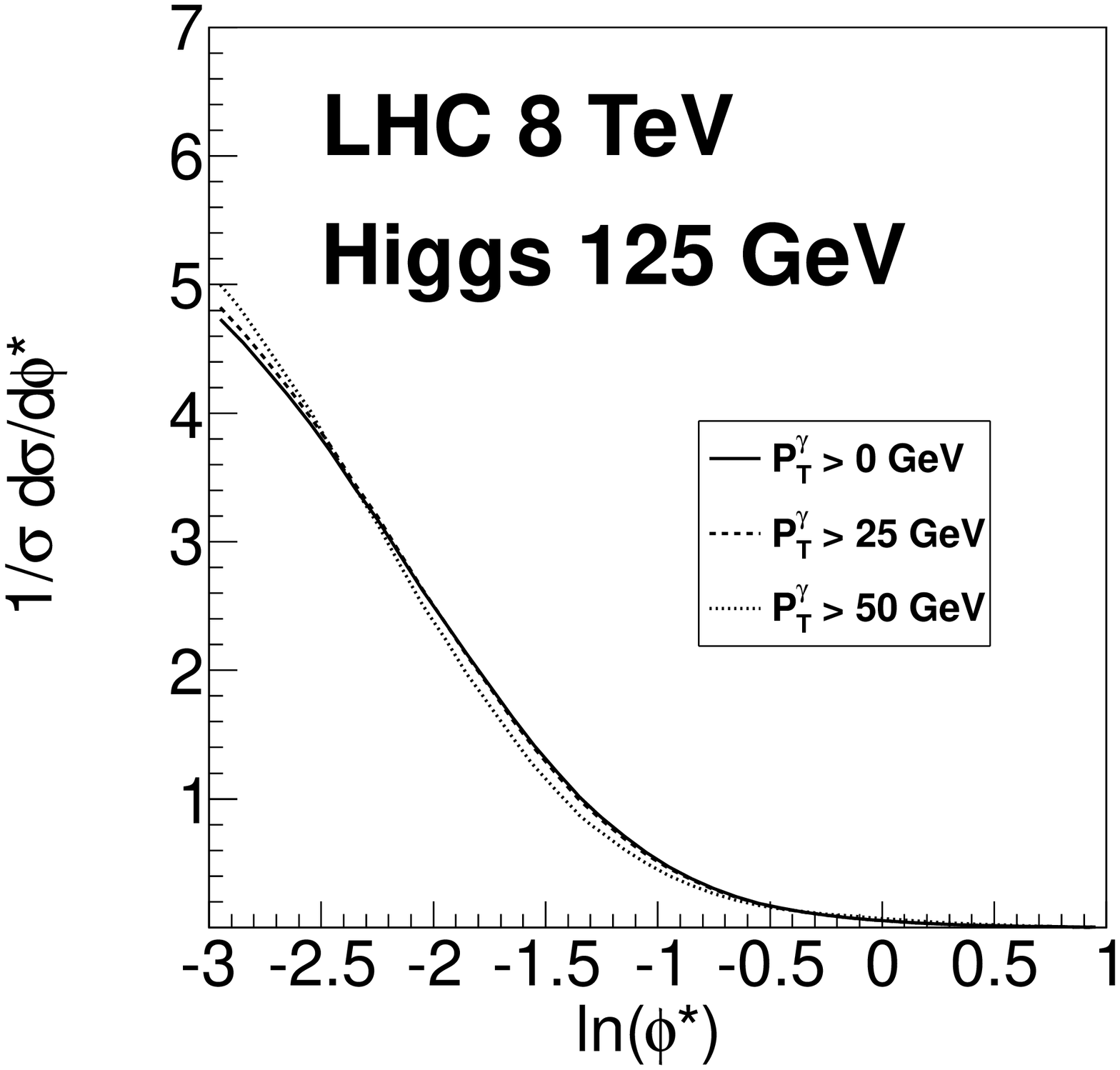}
\includegraphics[width=0.45\textwidth]{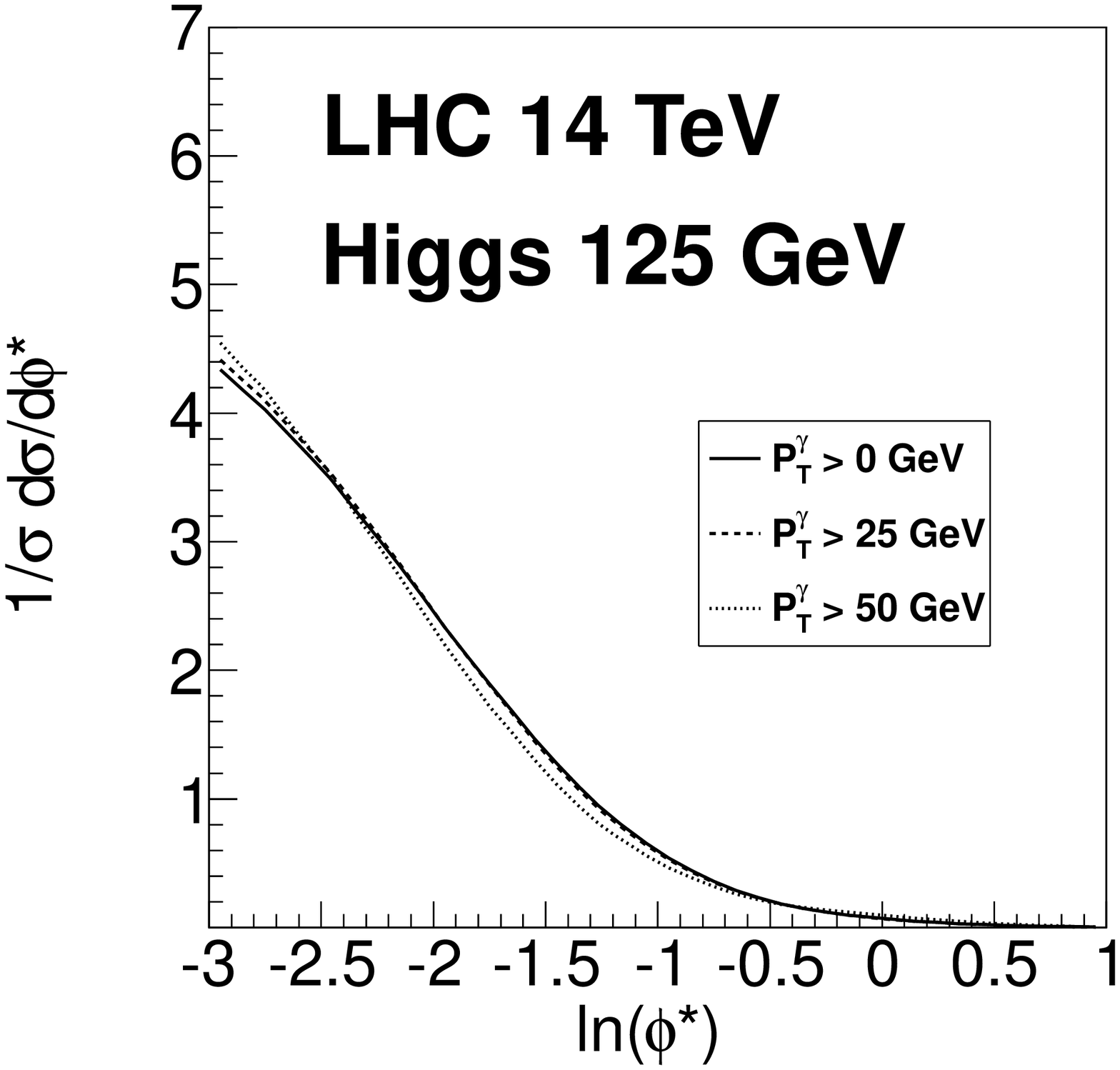}
\caption{The normalized $\phi^*$ distributions for different transverse momentum  cuts ($P_T^{\gamma}>0, 25, 50 $ GeV) on the individual photon
at the Tevatron (1.96 TeV)
and the LHC (7 TeV, 8 TeV and 14 TeV).}
\label{phi-cut}
\end{figure}

\begin{figure}[!htb]
\includegraphics[width=0.45\textwidth]{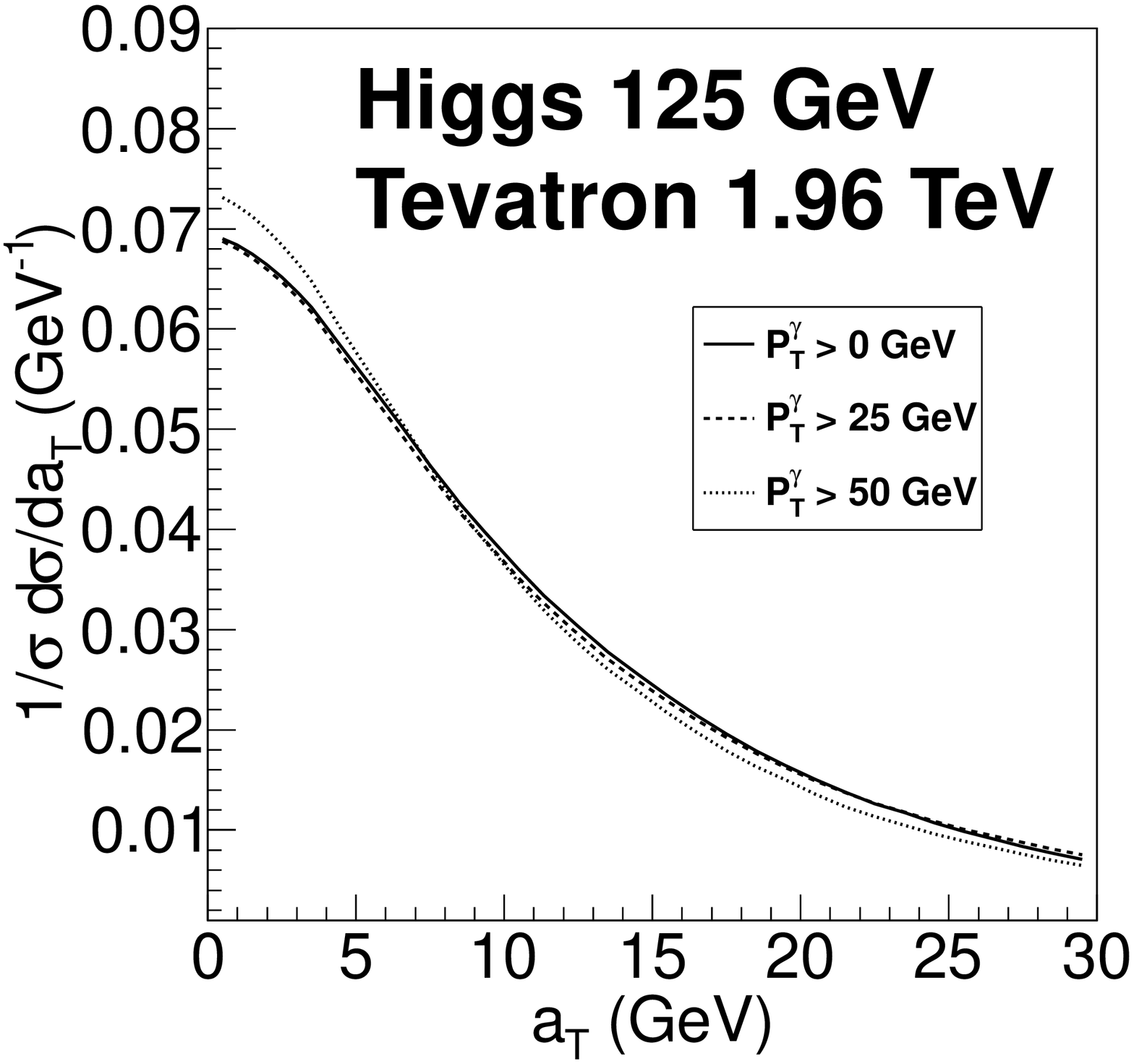}
\includegraphics[width=0.45\textwidth]{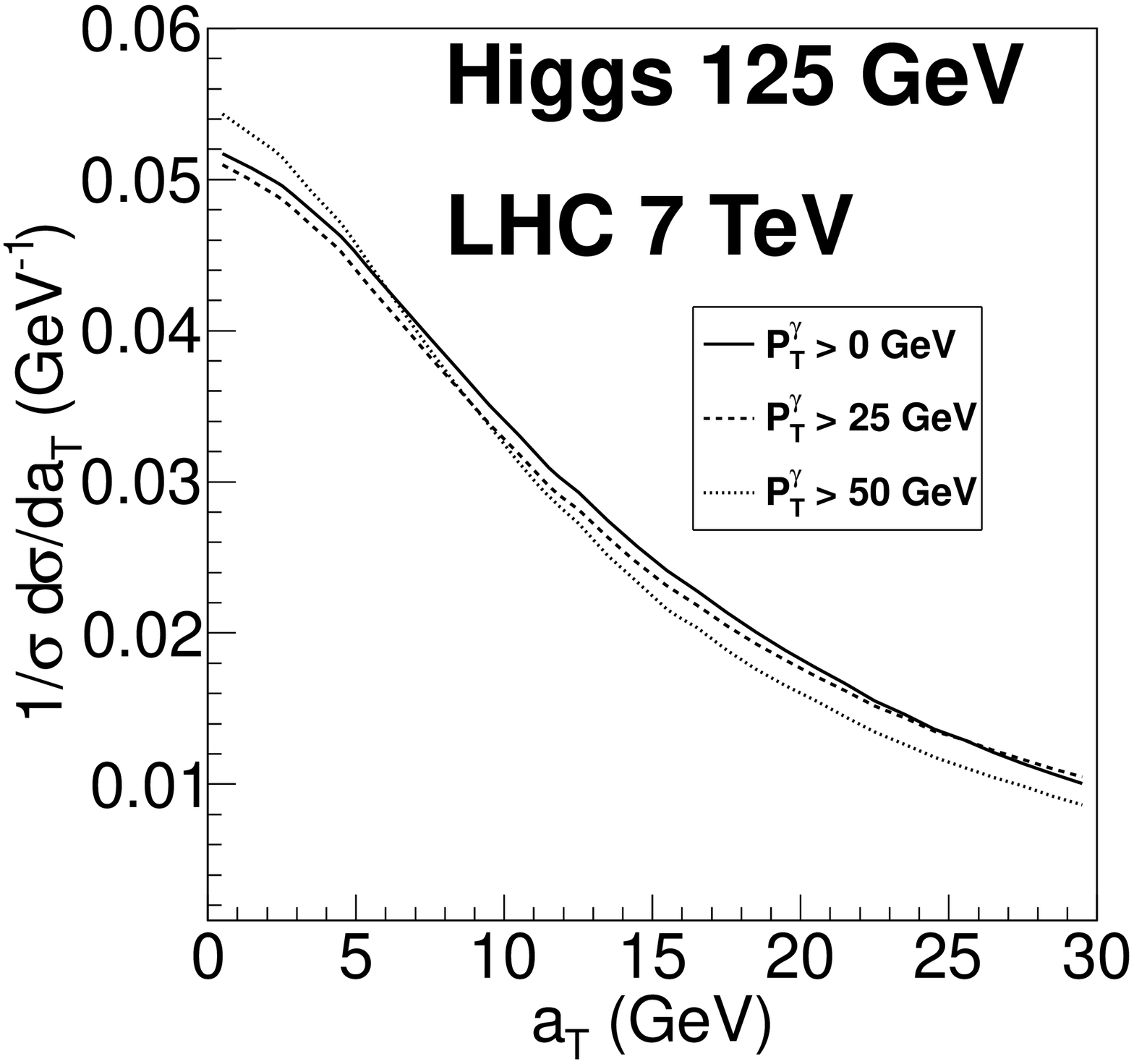}
\includegraphics[width=0.45\textwidth]{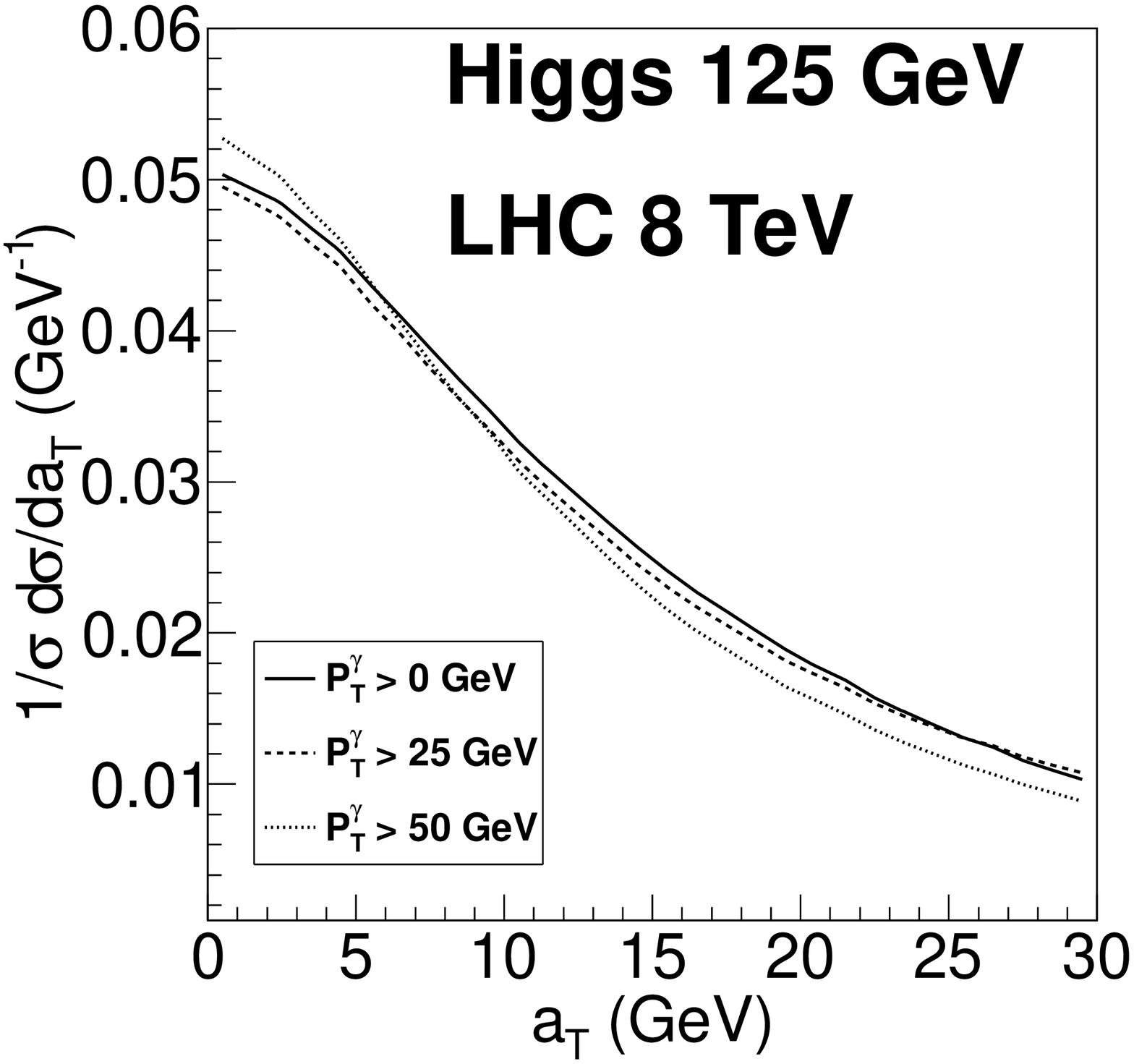}
\includegraphics[width=0.45\textwidth]{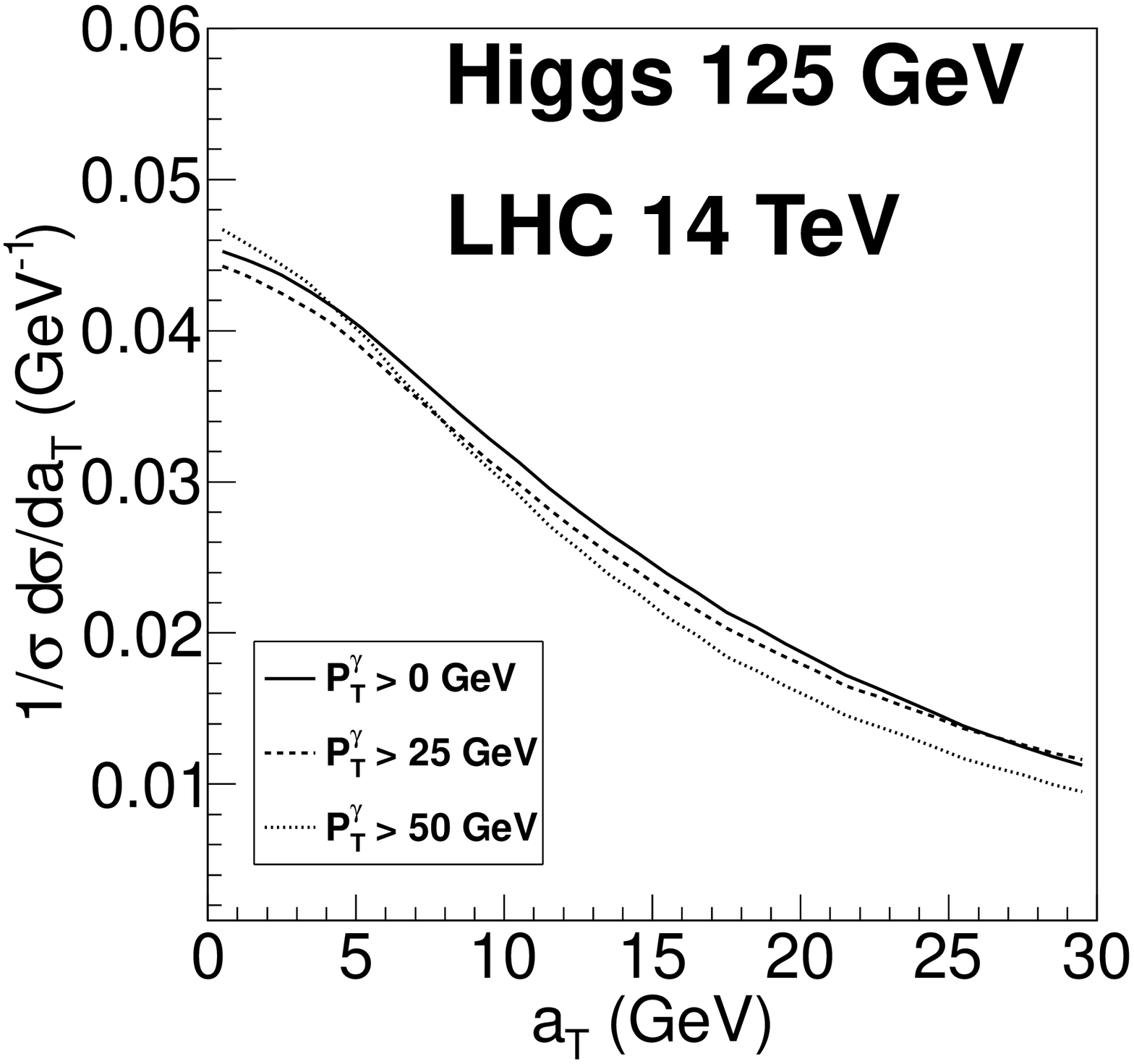}
\caption{The normalized $a_T$ distributions for different transverse momentum cuts ($P_T^{\gamma}>0, 25, 50 $ GeV)  on the individual photon
at the Tevatron (1.96 TeV)
and the LHC (7 TeV, 8 TeV and 14 TeV).}
\label{at-cut}
\end{figure}

\begin{figure}[!htb]
\includegraphics[width=0.45\textwidth]{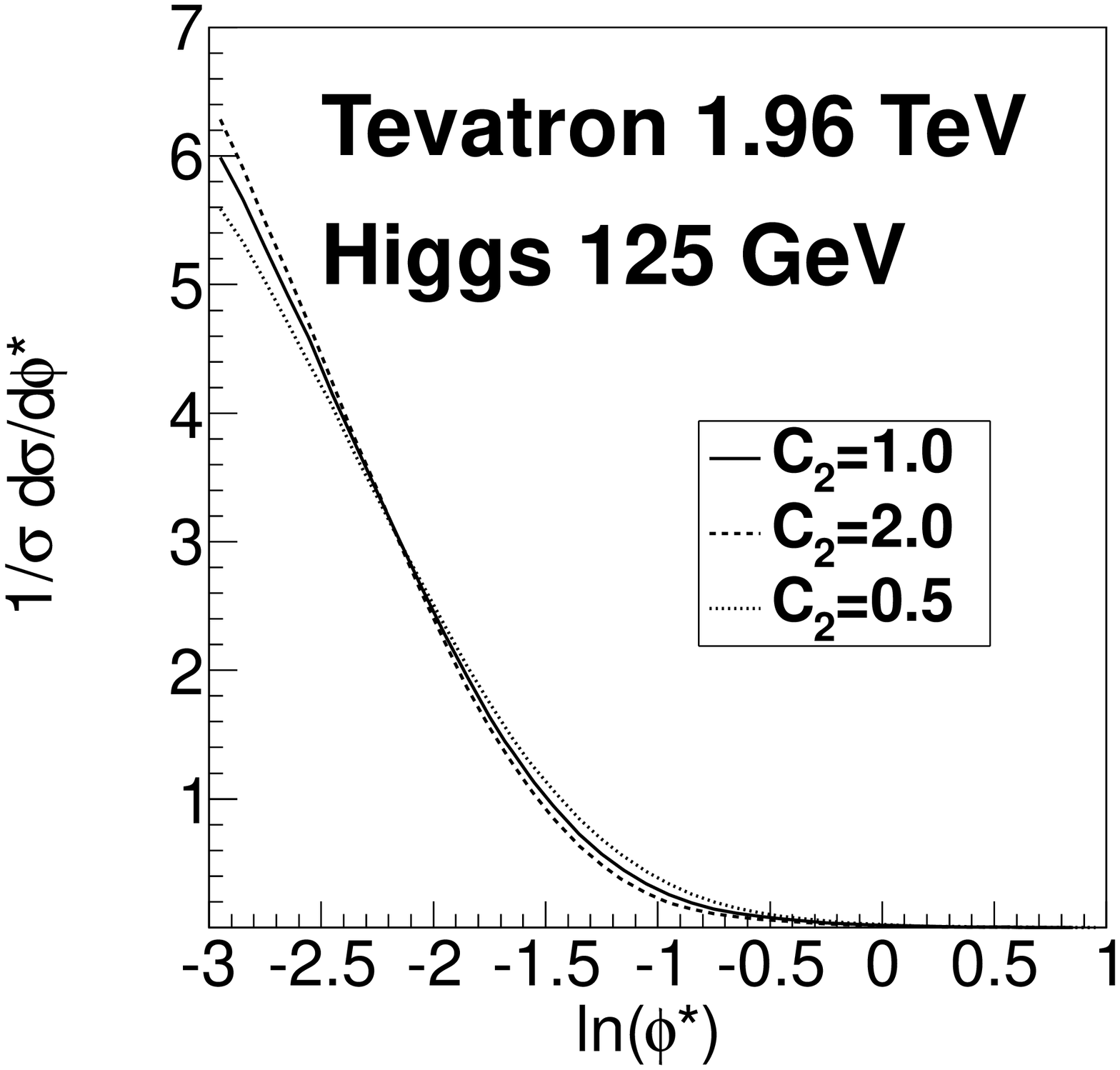}
\includegraphics[width=0.45\textwidth]{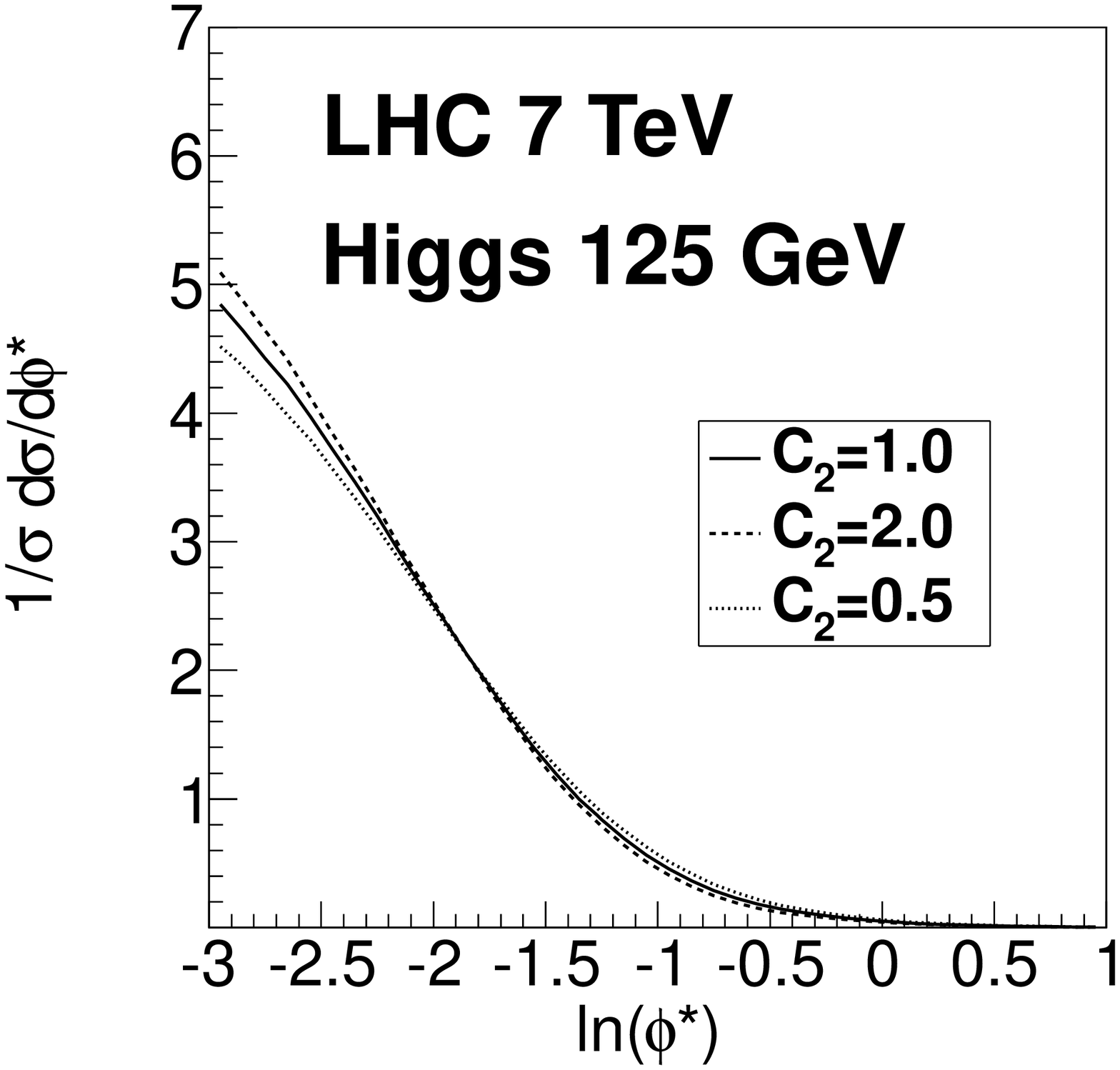}
\includegraphics[width=0.45\textwidth]{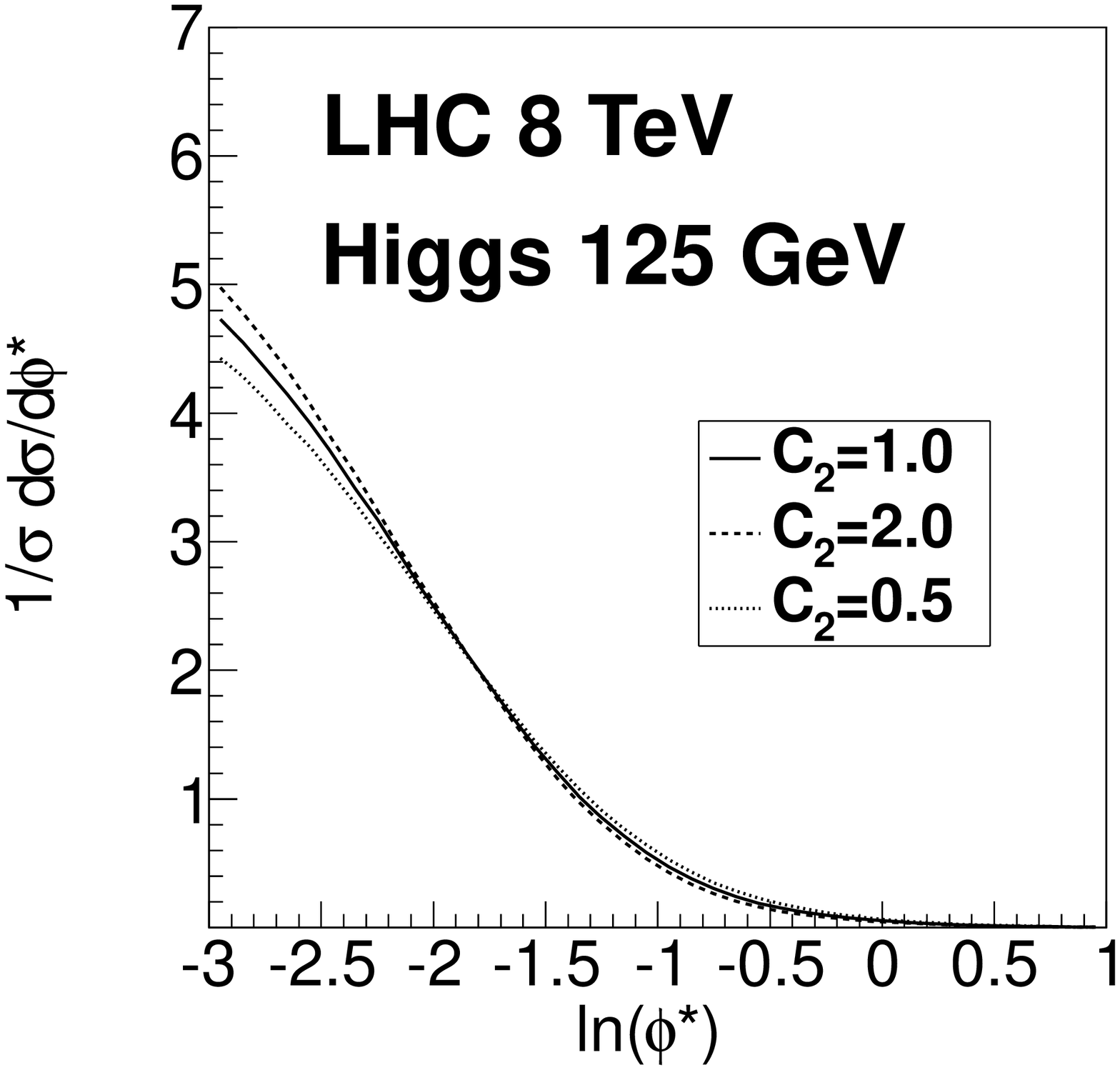}
\includegraphics[width=0.45\textwidth]{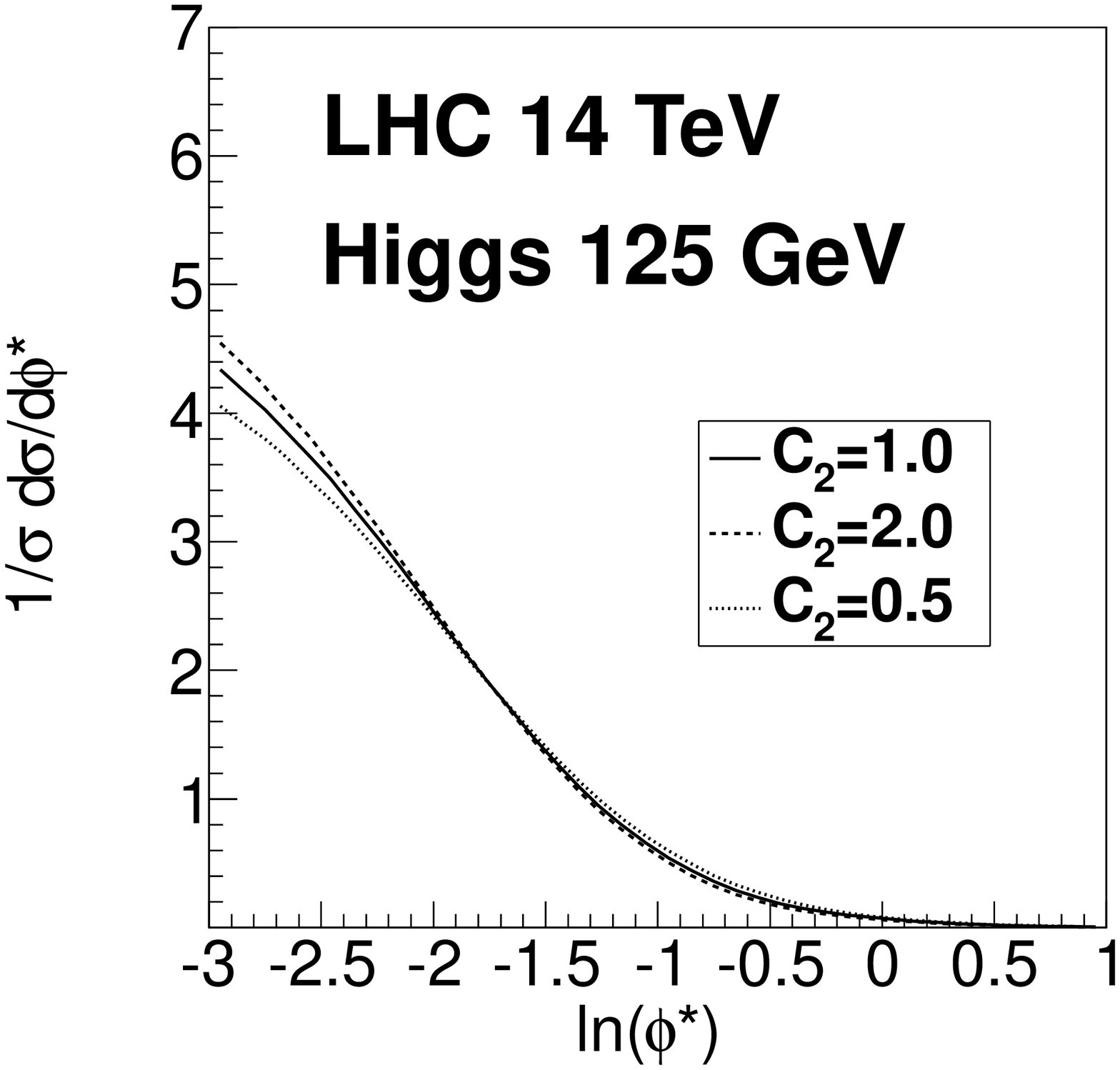}
\caption{The normalized $\phi^*$ distributions for different scale choices ($C_2=1,2,0.5$) on the individual photon
at the Tevatron (1.96 TeV)
and the LHC (7 TeV, 8 TeV and 14 TeV).}
\label{phi-scale}
\end{figure}

\begin{figure}[!htb]
\includegraphics[width=0.45\textwidth]{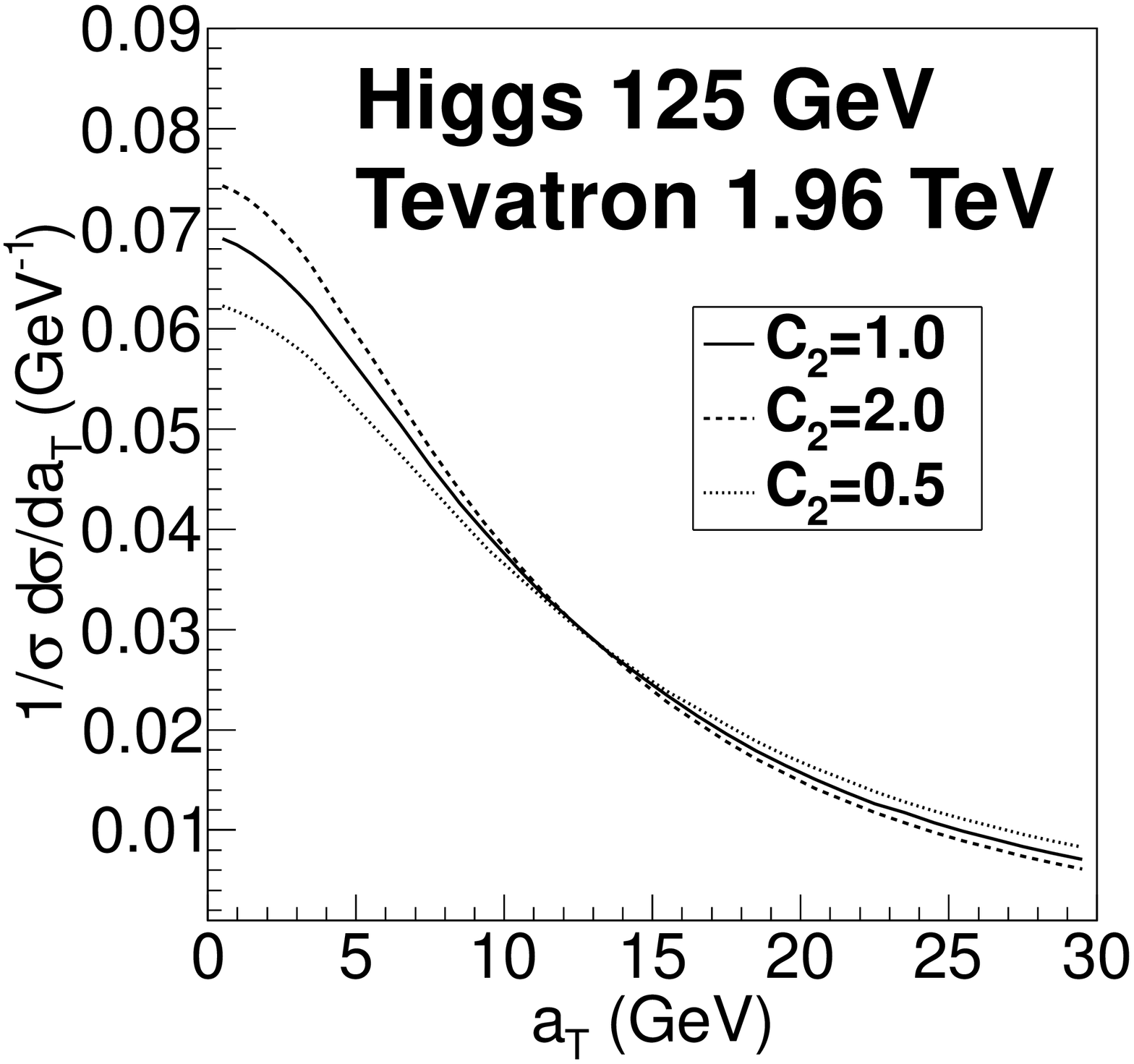}
\includegraphics[width=0.45\textwidth]{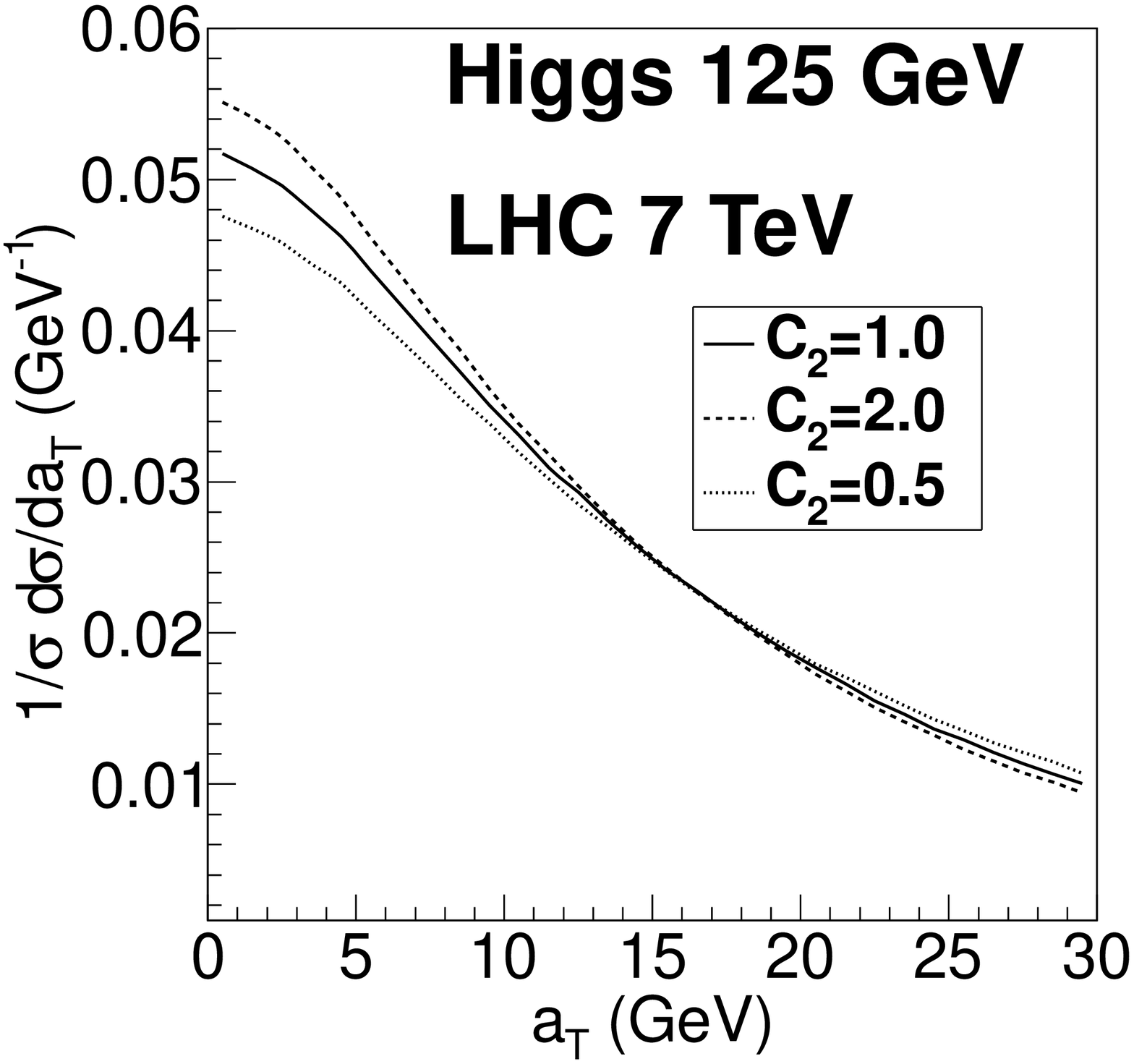}
\includegraphics[width=0.45\textwidth]{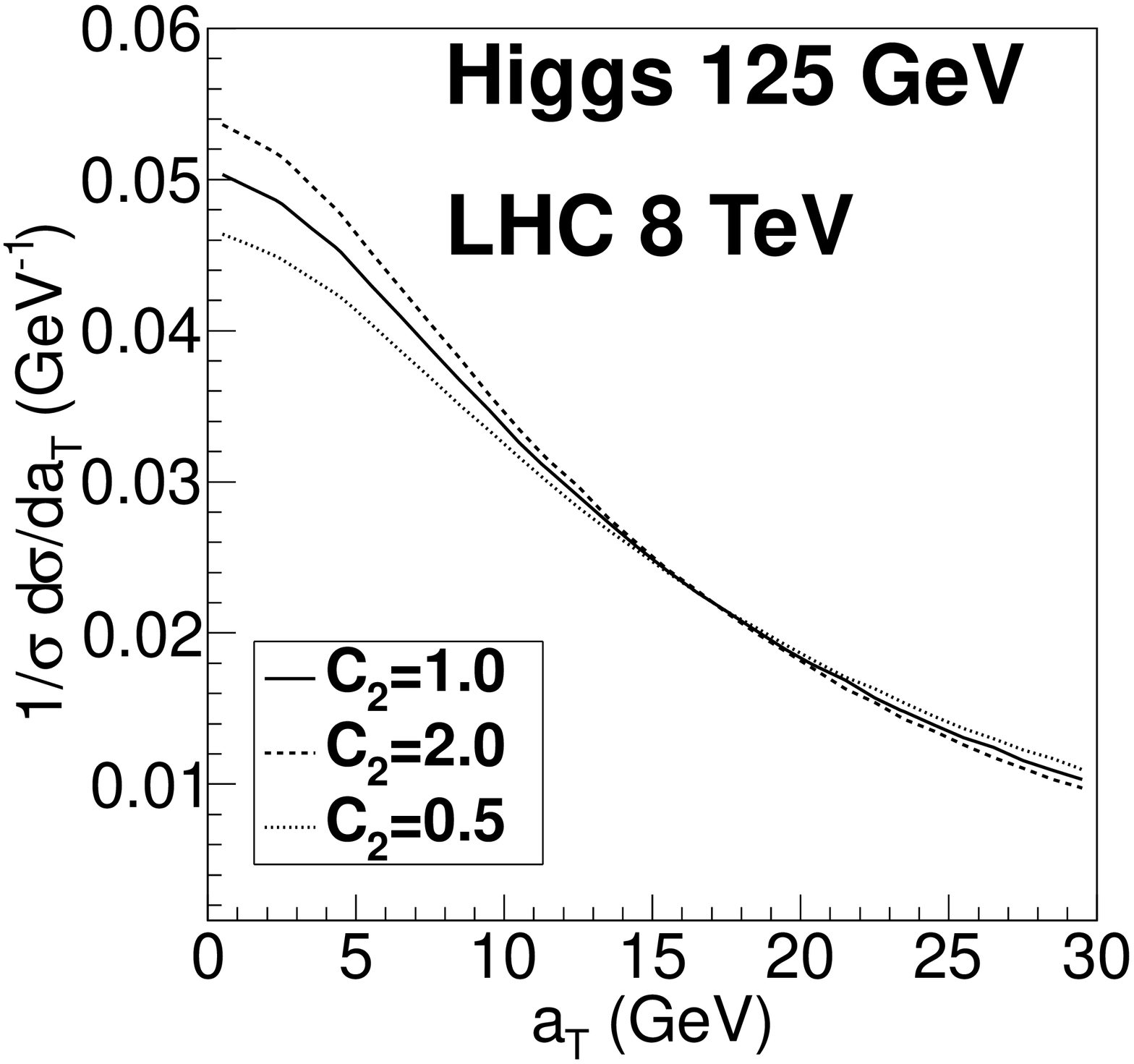}
\includegraphics[width=0.45\textwidth]{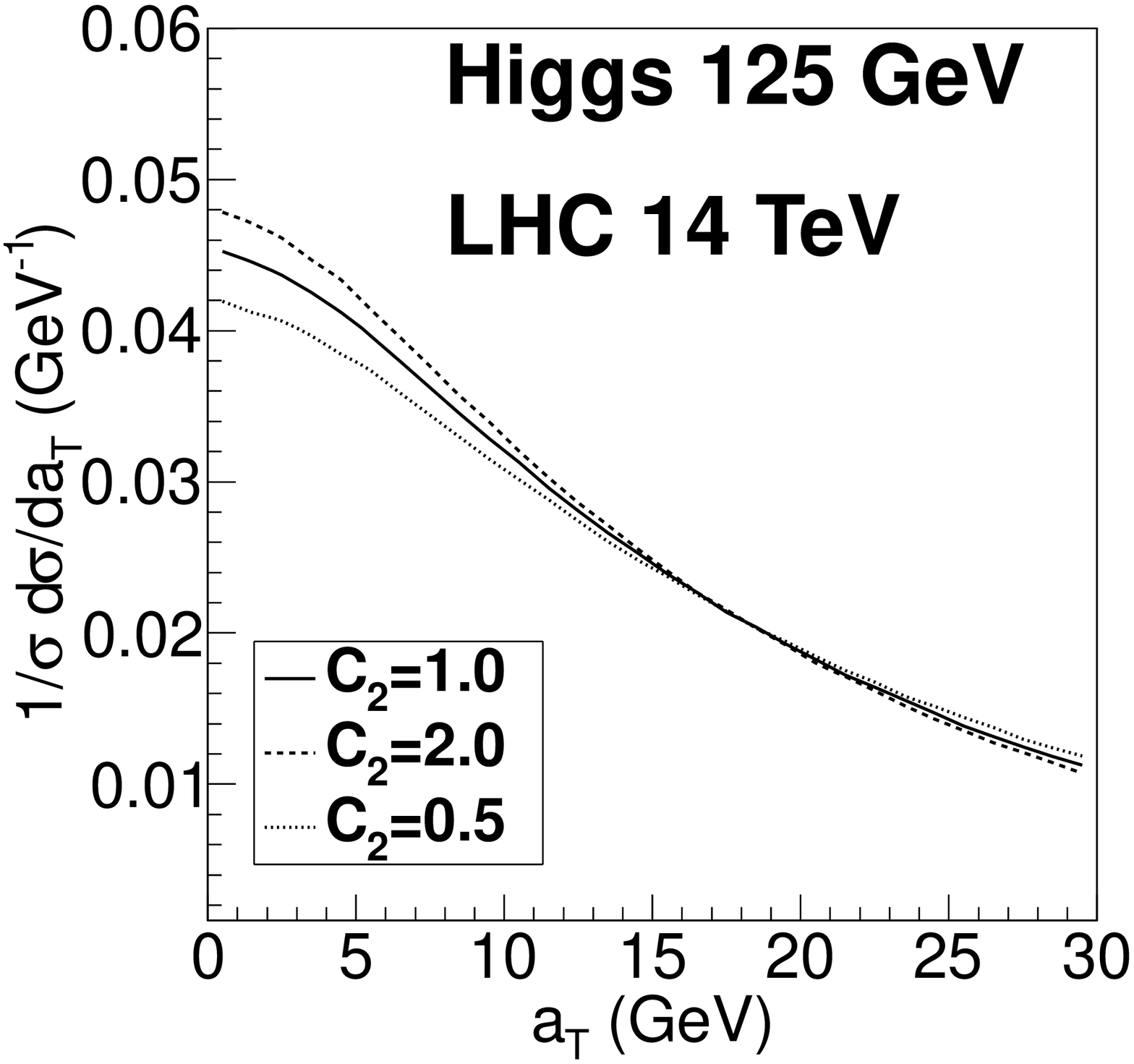}
\caption{The normalized $a_T$ distributions for different  scale choices ($C_2=1,2,0.5$)  on the individual photon
at the Tevatron (1.96 TeV)
and the LHC (7 TeV, 8 TeV and 14 TeV).}
\label{at-scale}
\end{figure}

Before closing this section, we would like to discuss the prediction of ResBos2 program
on the kinematical distributions of the two photons from the decay of Higgs boson produced at the Tevatron and the LHC.
When measuring the transverse momentum distribution of the $Z$ boson produced at the Tevatron,
the experimentalists have proposed two other variables, i.e. $a_T$ \cite{Vesterinen:2008hx} and $\phi^{*}$ \cite{Banfi:2010cf},
which have been shown to be less susceptible to the effects of experimental resolutions and efficiencies.
For example, the D\O~ collaboration at the Tevatron has measured the distribution of $\phi^{*}$
for the  $Z$ boson production and its decay into two leptons \cite{Abazov:2010mk}.
In the process $gg\to H\to \gamma\gamma$,
it is convenient to measure these two variables, similar to what was done in the analysis of the Drell-Yan lepton pair production.
Without restating their definitions given in Refs. \cite{Vesterinen:2008hx,Banfi:2010cf},
we show the normalized distributions of $\phi^{*}$  and $a_T$ for different transverse momentum cuts
on the individual photon in Fig. \ref{phi-cut} and Fig. \ref{at-cut}, respectively.
It is evident that the distributions of  $\phi^{*}$ are almost insensitive to the transverse momentum cuts
on the individual photon at both the Tevatron and the LHC.
From Fig. \ref{at-cut}, we can see that the distributions of $a_T$ change a little
when the transverse momentum cut on the individual photon varies from 0 GeV to 50 GeV.
Figure \ref{phi-scale} and Figure \ref{at-scale} show the scale dependencies of the distributions of $\phi^{*}$  and $a_T$, respectively.
It can be seen that in the small value region,
the scale uncertainties of the distributions of $\phi^{*}$ and $a_T$ is about $\pm 5\%$ and $\pm 8\%$ at most, respectively.
In the large value region, the scale uncertainties are negligibly small.

\section{Conclusion}
\label{sec:conc}
We  have improved the resummation calculations in the ResBos program for the Higgs boson production
via gluon-gluon fusion by including the NNLO Wilson coefficient functions and G-functions.
We show that including NNLO Wilson coefficient functions increases the total cross section predictions of ResBos
for a 125 GeV Higgs Boson production by about $8\%$ and  $6\%$ at the Tevatron and the LHC, respectively.
The predictions of the newly improved ResBos progran, dubbed ResBos2,
agree with the predictions of HNNLO and HqT2 programs within a couple of percent
at the Tevatron and the LHC.
We also show that for the transverse momentum distributions,
the predictions of ResBos2 generally increase the overall size
of the distributions in the low to medium transverse momentum regions, as compare to the old ResBos program.
Especially, the improvement in the small transverse momentum region can be as large as $20\%$
at both the Tevatron and the LHC,
and the peak positions are slightly shifted toward lower values than ResBos.
At the LHC, ResBos2 predictions for the transverse momentum distribution are almost the same as the HqT2 program.
At the Tevatron, the peak height of ResBos2 prediction is somewhat higher than that of the HqT2 program.
When increasing the renormalization constant $C_2$ and keeping the relations between $C_{i}, i=1,2,3,4$,
in the ResBos2 calculations,
the distributions in the low and high $Q_T$ regions are increased and suppressed, respectively.
The inclusion of G-functions in ResBos2 can mainly slightly modify
the shape of the transverse momentum distribution of the Higgs boson at the LHC, by less than $1\%$.
Similarly, varying the non-perturbative coefficients by a factor of two
only slightly modifies the peak position and the height of the $Q_T$ distribution.
Thus, the transverse momentum distributions of the Higgs boson produced at the LHC
are dominated by the perturbative Sudakov resummation effect of multiple soft gluon emissions,
and the effects of non-perturbative physics are not as important.

From an experimental viewpoint, we also present the distributions of two variables $a_T$  and $\phi^{*}$
in the process $gg\to H \to \gamma\gamma$,
which have better experimental resolutions than $Q_T$ itself.
We find that the distributions of  $\phi^{*}$ are almost insensitive to
the transverse momentum cuts on the individual photon at both the Tevatron and the LHC.
The scale uncertainties of the distributions of $\phi^{*}$  in the small value region are about $\pm 5\%$ at most,
and the scale uncertainties in the large value region are negligible.
Similar conclusions also hold for the $a_T$ distributions.

\acknowledgments{
This work was partially supported by the National Natural Science Foundation of China,
under Grants No. 11021092, No. 10975004 and No. 11135003,
and by the U.S. National Science Foundation under Grant No. PHY-0855561.
}
\newpage
\newpage
\newpage

\bibliography{higgsc2}

\end{document}